\documentclass[letterpaper,twocolumn,10pt]{article}
\usepackage[accepted]{usenix}

\usepackage{tikz}

\usepackage{filecontents}
\usepackage[utf8]{inputenc} 
\usepackage[T1]{fontenc}    
\usepackage{hyperref}       
\usepackage{url}            
\usepackage{booktabs}       
\usepackage{amsfonts}       
\usepackage{nicefrac}       
\usepackage{microtype}      
\usepackage{xcolor}         
\usepackage{booktabs}
\usepackage{subcaption}
\usepackage{multirow}
\usepackage{amssymb}  
\usepackage{booktabs}
\usepackage{algorithm}
\usepackage{algorithmic}
\usepackage{placeins}   
\usepackage{natbib}
\usepackage{graphicx}

\usepackage{amsmath}

\usepackage[most]{tcolorbox} 

\usepackage{amsmath,amsfonts,bm}

\def\eqref#1{equation~\ref{#1}}

\def\1{\bm{1}}

\DeclareMathAlphabet{\mathsfit}{\encodingdefault}{\sfdefault}{m}{sl}
\SetMathAlphabet{\mathsfit}{bold}{\encodingdefault}{\sfdefault}{bx}{n}

\usepackage{enumitem}

\usepackage{url}
\usepackage{cleveref}
\definecolor{lightblue}{HTML}{D9F4FF}
\definecolor{lightred}{HTML}{FAD4D5}
\definecolor{mydarkblue}{rgb}{0,0.08,0.45}
\hypersetup{
  colorlinks=true,
  linkcolor=mydarkblue,
  citecolor=mydarkblue,
  urlcolor=mydarkblue
}

\icmltitlerunning{Aggressive Compression Enables LLM Weight Theft}

\begin{document}

\twocolumn[
\icmltitle{Aggressive Compression Enables LLM Weight Theft}

\begin{icmlauthorlist}
\icmlauthor{Davis Brown}{upenn}
\icmlauthor{Juan-Pablo Rivera}{gatech}
\icmlauthor{Dan Hendrycks}{cais}
\icmlauthor{Mantas Mazeika}{cais}
\end{icmlauthorlist}

\icmlaffiliation{upenn}{University of Pennsylvania, Philadelphia, PA, USA}
\icmlaffiliation{gatech}{Georgia Institute of Technology, Atlanta, GA, USA}
\icmlaffiliation{cais}{Center for AI Safety, San Francisco, CA, USA}

\icmlcorrespondingauthor{Davis Brown}{davisrbr@seas.upenn.edu}

\icmlkeywords{Machine Learning, ICML}

\vskip 0.3in
]

\printAffiliationsAndNotice{}

\begin{abstract}

As frontier AIs become more powerful and costly to develop, adversaries have increasing incentives to steal model weights by mounting exfiltration attacks. 
In this work, we consider exfiltration attacks where an adversary attempts to sneak model weights out of a datacenter over a network.
While exfiltration attacks are multi-step cyber attacks, we demonstrate that a single factor, the \textit{compressibility} of model weights, significantly heightens exfiltration risk for large language models (LLMs). 
We tailor compression specifically for exfiltration by relaxing decompression constraints and demonstrate that attackers could achieve $16\times$ to $100\times$ compression with minimal trade-offs, reducing the time it would take for an attacker to illicitly transmit model weights from the defender's server from months to days. Finally, we study defenses designed to reduce exfiltration risk in three distinct ways--making models harder to compress, making them harder to `find,' and tracking provenance for post-attack analysis using forensic watermarks. While all defenses are promising, the forensic watermark defense is both effective and cheap, and therefore is a particularly attractive lever for mitigating weight-exfiltration risk. 
\end{abstract}

\section{Introduction}

The cost of training frontier AI models has skyrocketed, with each new generation of models requiring exponentially more compute to train \citep{cottier2024rising}. Beyond their economic significance, advanced AI systems are increasingly viewed as critical assets in national security, given their rapidly improving capabilities. 
This dual economic and strategic importance has led to increased interest in securing model weights from theft \citep{nevo2024securing}. 

\paragraph{Weight exfiltration attacks.}
Of particular concern are \textit{weight exfiltration} attacks, where the data center hosting the model weights is compromised by an attacker.
This allows the attacker to steal the weights of the language model by smuggling them out over the network.
However, the risk of weight exfiltration attacks remains poorly understood, with much uncertainty surrounding their feasibility.

\begin{figure*}[t]
    \centering
     \includegraphics[width=\textwidth]{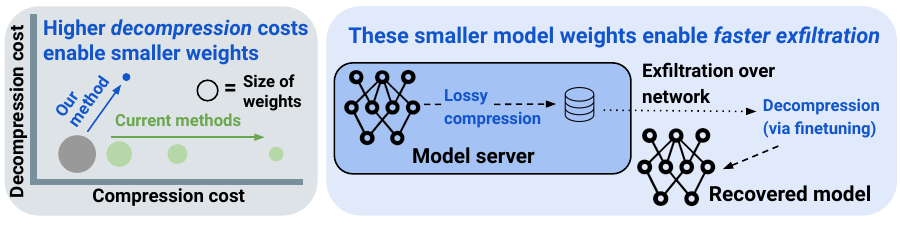}
    \caption{Prior compression methods are optimized for efficient inference, with minimal decompression overhead. By relaxing these constraints, we show that much smaller model weights can be achieved. This is highly relevant for weight exfiltration, as it reduces both exfiltration time and detection risk. After being stolen from the server, the model can be decompressed and fine-tuned to recover performance at far lower cost than training from scratch.}
    \label{fig:compression-decompression}
    \vspace{-10pt}
\end{figure*}

A common tactic in standard data exfiltration attacks is to compress data before transmission over the network \citep{mitre_t1560_001}, reducing the likelihood of detection. However, this tactic has not been extensively studied in the context of weight exfiltration attacks. While prior work on large language model (LLM) compression has focused on optimizing models for efficient inference---achieving high fidelity with up to $4\times$ compression \citep{liu2024spinquant}---the requirements for inference differ significantly from those in weight exfiltration scenarios. In particular, existing methods are designed to have an efficient forward pass after the initial compression--- we refer to the cost of obtain useful weights after they have been compressed as the `decompression cost` in \Cref{fig:compression-decompression}. These differences suggest that more aggressive lossy compression techniques, tailored for exfiltration, could be feasible, leading to a higher risk of successful attacks.

To study how to better defend language model weights from being stolen from servers, we investigate the impact of LLM compression on the feasibility of weight exfiltration attacks. We begin by proposing a simple quantitative model for exfiltration success, which is influenced by factors such as the compression ratio. Next, we show that compression techniques specifically optimized for exfiltration can significantly reduce exfiltration time and improve the likelihood of a successful attack. By relaxing the typical constraints for model compression that prioritize still being able to run the model efficiently for inference, we achieve substantially higher compression rates than existing methods, reaching well over $16\times$ compression with minimal trade-offs. 


Finally, we consider three candidate `model-level' defenses in detail.
These defenses attempt to harden weights against an adversary in three ways: by making them harder to compress, via a moving target defense first proposed in \citep{shlegeris2023gauge}, and by forensic watermarking.
We find that forensic watermarking is the most promising--- it is both cheap to implement and reasonably robust.

In summary, we make the following contributions.
\begin{itemize}
    \item \textbf{Defining a new threat model.} We build a model to quantify the risk of weight exfiltration under various conditions and identify \emph{model compressibility} as a crucial neglected factor. Indeed, model compressibility may make exfiltration attacks much more feasible.
    \item \textbf{Identifying new phenomena around extreme compression.} We find that models are orders of magnitude (10-100x) more compressible than standard practice would suggest, when one allows for high \textit{decompression} costs, in the form of additional finetuning. While high decompression costs are not feasible for model inference quantization, high costs are acceptable for attackers while stealing model weights.
    \item \textbf{Proposing reasonable defenses.} We review and measure a few baseline model- and system-level defenses, and find that \textit{forensic watermarking} of layer weights for the purpose of provenance is particularly appealing.
\end{itemize}
Our results suggest that LLM compression is a significant factor affecting the risk of weight exfiltration attacks. Attacks that would have taken months before may be feasible in days with advanced compression techniques. This motivates further research on the risk of weight exfiltration and further investment in securing model weights. 



\section{Related Work}

\paragraph{Model Stealing.}
Black-box model stealing attacks have demonstrated that an adversary with only API access to a machine learning service can steal a deployed model’s functionality.
For example, \citep{10.5555/3241094.3241142} extracts the exact weights of a logistic classifier with only API access and \citep{carlini2024stealing} steals the exact final layer of a model without training. 
Another line of work attempts to distill the \textit{capabilities} of a model with query-only access \citep{shridhar2023distilling,DBLP:conf/iclr/Gu0WH24,li2024distilling}, however, the resulting models typically significantly underperform the teacher model.

\paragraph{Data exfiltration.}
A long line of work in the security literature has studied data exfiltration threats. Particularly relevant to our setting are advanced persistent threats (APTs) \citep{alshamrani2019survey}. These attacks involve targeted infiltration and long dwell times on compromised servers to gather and slowly exfiltrate high-value data. In these settings, the amount of data to exfiltrate is a key consideration, and compression techniques are sometimes used \citep{mitre_t1560_001}. Several defenses against APTs have been studied. Moving target defenses periodically modify aspects of the infiltrated system to impede adversaries \citep{crouse2015probabilistic, sengupta2020survey}. In the context of LLM exfiltration, \citet{greenblatt2024preventing} propose enforcing minimal upload limits on LLM servers to meet customer demand.
Similarly, \citep{rinberg2025verifying} study a defense mechanism (inference verification) that makes steganography more difficult, assuming similar upload limits.

\paragraph{LLM quantization and compression.}
A key factor in weight exfiltration attacks is the size of the weights to steal, which can be reduced through compression. \citet{nevo2024securing} note that there is much uncertainty around the effective size of model weights, with models commonly served at $2\times$ or even $4\times$ compression. This is enabled by recent work on model quantization for efficient inference, which achieves considerable performance through sophisticated quantization schemes \citep{zhu2023survey, frantar2022gptq, egiazarian2024extreme, liu2024spinquant, tseng2024quip}. However, research on model quantization has not previously considered the weight exfiltration setting. 
A key contribution of our work is to demonstrate that this setting enables \textit{achieving far stronger compression, increasing the risk of weight exfiltration}.
Similarly, to our knowledge we are the first work to consider a defense to make model weights more difficult to compress in order to defend against weight exfiltration attacks.

\paragraph{Model weight watermarking.} 
Methods to watermark neural network embed a proof of ownership into model parameters to help protect the intellectual property of model owners  \citep{uchida2017embedding, rouhani2018deepsigns, Wang2019AttacksOD, pagnotta2022tattooed0}. We consider the setting of watermarking for model weight provenance post-exfiltration.





\section{Weight Exfiltration}

Here, we describe the specific threat model that we consider. We base our quantitative model of attacker success on this threat model. 

\subsection{Threat Model}

While standard data exfiltration attacks have been extensively studied and reported on in the cybersecurity literature \citep{mitre_t1560_001}, frontier AI inference workloads are significantly different from standard network and application architectures. 
Notably, the attacker must send a significant amount of data (the model weights) out of the datacenter.
Compared to data exfiltration attacks to obtain e.g., individual customer data, this makes it harder on the attacker. We first formalize the objective and affordances of the attacker:

\begin{tcolorbox}[
    enhanced,
    colframe=red!80!black,
    title= \centering \textbf{ \large Attacker objective and affordances},
    titlerule=0.2mm,
    width=0.9\linewidth,
    center,
    rounded corners,
    before skip=6pt,
    after skip=6pt,
]
\normalsize
\textbf{Objective.} We consider a standard Advanced Persistent Threat (APT) with the goal of stealing model weights from an inference server that outputs text, image, video, or audio data to users. Note that the attacker is not attempting to distill the model from sampled inputs and outputs, but wants to exfiltrate the actual weights of the served language model.

\smallskip

\textbf{Affordances.} The attacker has already compromised the model server and gained access to unencrypted model weights. However, the attacker is not a physical insider, and thus cannot upload weights to a USB drive. They must instead extract the weights over the server's network. The attacker is restricted in their upload rate because they need to evade detection until the attack is complete. Additionally, the attacker must hide exfiltrated data within legitimate outputs (for example, via steganography), as the defender uses deep packet inspection to ensure that all outgoing traffic conforms to expected formats.
\smallskip

\end{tcolorbox}

Thus, we assume the attacker creates fake accounts and extracts information through responses generated by the model (a form of Sybil attack \citep{douceur2002sybil}). We assume that partial exfiltration of the model is far less valuable than full exfiltration. We do not consider hardware security measures. Likewise, the specifics of how the server was initially compromised are out of scope. Next, we consider the objective and affordances of the model provider:

\begin{tcolorbox}[
    enhanced,
    colframe=blue!80!black,
    title= \centering \textbf{ \large Defender objective and affordances},
    titlerule=0.2mm,
    width=0.9\linewidth,
    center,
    rounded corners,
    before skip=6pt,
    after skip=6pt,
]
\normalsize
\textbf{Objective.} The defender monitors server activity and network traffic for signs of APT activity. While detection is not guaranteed, prolonged adversary activity increases the likelihood that the defender will detect the attack attempt. We assume the defender has a fixed probability of detecting malicious behavior per unit time.

\smallskip

\textbf{Affordances.} The network bandwidth from the server is right-sized to the expected user traffic, and deep packet inspection is employed to ensure that all outgoing traffic matches expected formats. This forces the attacker to encode data via e.g. steganography.
\smallskip

\end{tcolorbox}

Note that this threat model is only one possible set of assumptions. We discuss limitations of this threat model in \Cref{app:limitations}.

\subsection{Quantitative Model of Weight Exfiltration}
\label{sec:quantitative_model}

\begin{figure*}[t]
    \centering
    \includegraphics[width=\textwidth]{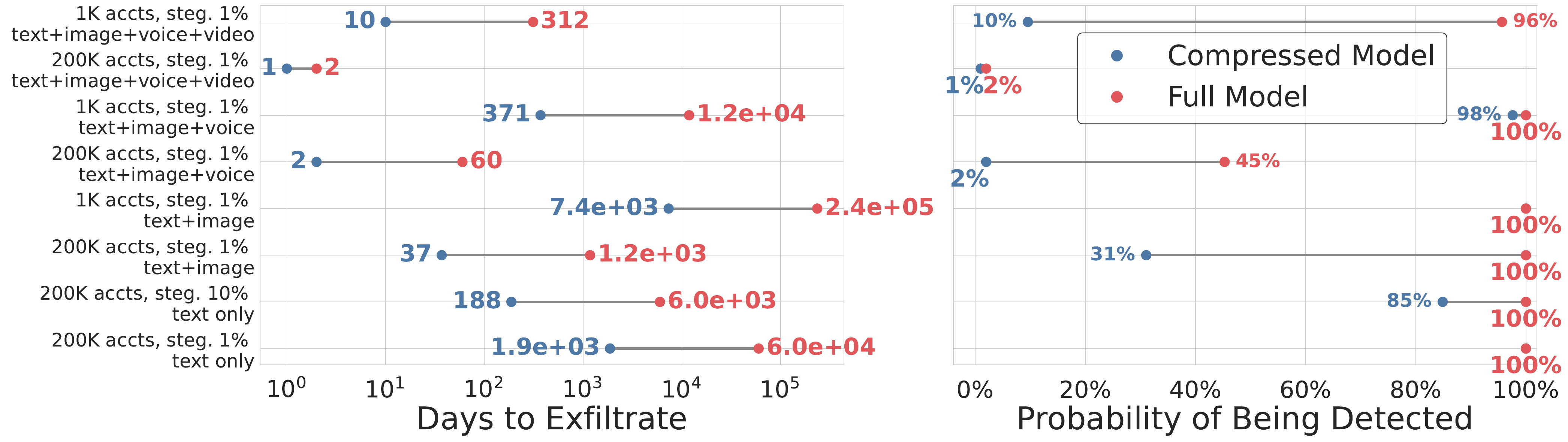}
    \caption{Model compression enables more successful weight exfiltration attacks. We compute the rate of weight exfiltration (left) and the probability of detection (right) for full models and models compressed using our method designed for the weight exfiltration setting. Probability of detection is computed from days to exfiltrate. For more details, see \Cref{sec:quantitative_model}.}
    \label{fig:exfil-dumbbell}
\end{figure*}



To demonstrate that smaller, compressed weights are easier to steal, we use a simple quantitative model roughly calibrated to the scenario of a weight exfiltration attack on GPT-4 \citep{achiam2023gpt}. Our model computes the time required for exfiltration and the probability of success given a fixed daily probability of detection. We say attack fails if it is detected before the model is fully exfiltrated, in keeping with our threat model. Our model incorporates various factors and three different usage profiles, allowing us to study a wide variety of attack scenarios. We describe these below.

\paragraph{Factors.} Our model starts from the assumption that the exfiltration occurs on an inference server for a public API. The API has daily users, some number of which are compromised or created by the attacker. The attacker then attempts to exfiltrate the weights over the network, while avoiding detection. We describe the factors in our model below.

\begin{itemize}[leftmargin=*]
    \item \textbf{Network egress per user per day} ($E$): The number of bits sent to each user of the inference API per day.
    \item \textbf{Number of compromised daily users} ($N$): We assume the attacker mounts a Sybil attack \citep{douceur2002sybil} and can only receive data through steganography on legitimate channels. Note that the attacker may have more accounts than they use each day, swapping out accounts to mimic realistic usage patterns.
    \item \textbf{Steganography ratio} ($s$): This is the fraction of output bits that the attacker can use for hiding arbitrary data, used for exfiltrating the model weights.
    \item \textbf{Model size} ($M$): The size of the model to be extracted in bits, before compression is applied.
    \item \textbf{Compression ratio} ($c$): The multiplicative reduction in model size after applying compression.
    \item \textbf{Probability of detection per day} ($p$): The probability that the attacker's activity is detected each day. For simplicity, we assume this is a fixed number.
\end{itemize}

Given these factors, we compute the time until exfiltration in days as
\[T = \frac{M \times c}{E \times N \times s}.\]
The probability of successful exfiltration is computed as $(1 - p)^T$.

\paragraph{Usage profiles.}
We consider four possibilities for the average daily user of an LLM inference API. These correspond to current and anticipated future modes of interaction with LLMs. In all cases, we assume that outputs are converted to standard media formats before being sent over the network to users. These usage profiles are described below.

\begin{itemize}[leftmargin=*]
    \item \textbf{Text Only:} The average user in this scenario generates $10$ pages of text per day ($500$ English words per page). This is in the ballpark of current daily average usage of models such as GPT-4 \citep{Altman2024}. In reality, some users may generate far more or far less. We assume all users have equal consumption for simplicity. 
    \item \textbf{Text + Images:} In addition to the above text usage, users in this scenario generate $1$ image per day on average.
    \item \textbf{Text + Images + Voice:} In addition to the usage of the previous scenario, the average user in this scenario generates $30$ minutes of voice/audio output per day. This may be used for talking with an LLM as an alternative mode of communication.
    \item \textbf{Text + Images + Voice + Video:} This is a hypothetical future mode of interaction with LLMs involving real-time video output, possibly for interaction with chatbot avatars.
\end{itemize}

In Appendix D, we derive estimates for the factors in our model and for different usage profiles. In some cases, we consider ranges of values for a given factor.
\section{Compression Experiments}

Here, we describe our experiments demonstrating that compression techniques can be tailored to the weight exfiltration setting, enabling improvements to the compression ratio. We then evaluate how this affects exfiltration time and success rates under our quantitative model of exfiltration.

\begin{figure*}[t]
    \vspace{-10pt}
    \centering
    \includegraphics[width=0.9\textwidth]{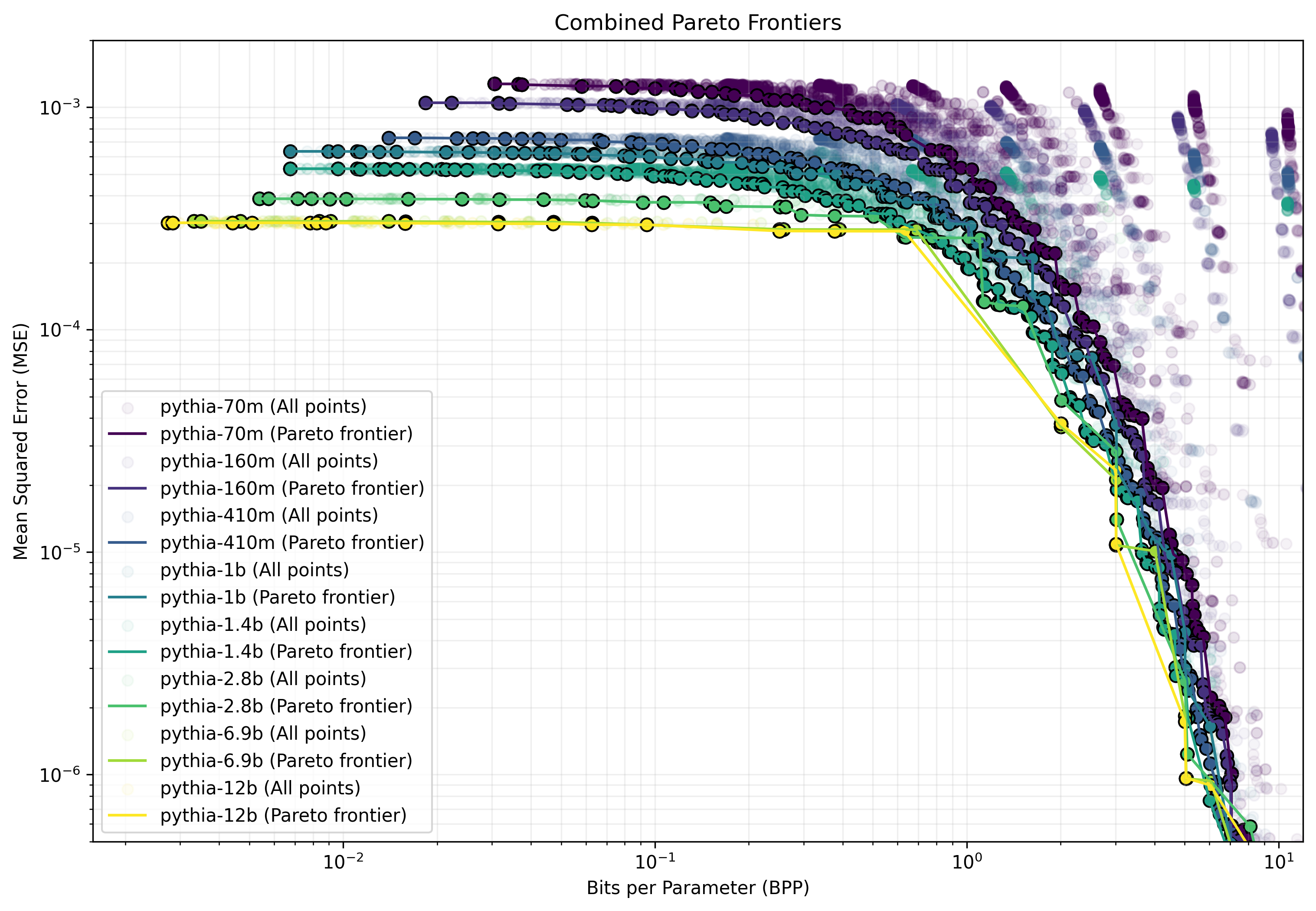}
    \caption{The MSE between the original and compressed weights vs the bits-per-parameter (BPP), along with the Pareto frontier for MSE vs BPP, for each Pythia model. We observe that in weight space, larger models are easier to compress.}
    \label{fig:compression-comparison-weights}
    \vspace{-10pt}
\end{figure*}

\subsection{Rate-Distortion in Weight Space}
\label{app:rate_distortion}

We find that model weights present a smooth tradeoff between how much they are compressed, measured in bits-per-parameter, and the fidelity of their compression, measured in MSE (mean square error between the original weights and the compressed weights). We sample from the Pareto frontier of the compression hyperparameters for our method, showing the results in \Cref{fig:compression-comparison-weights}, and \Cref{fig:compression-comparison-trends,fig:kmeans-comparison} in the appendix. For these experiments, we use a broader range of models than in the main experiments, including the Pythia suite \citep{biderman2023pythia} and Llama 3 models \citep{dubey2024llama}. These results reveal several interesting phenomena, which we describe in the following paragraphs.

\paragraph{Rate-distortion curves reveal smooth trade-offs.}
Our main results in \Cref{tab:compression-comparison-combined} show that models can be compressed using our method with very little degradation in downstream metrics like log-loss. To obtain a more fine-grained view of how our method performs at different BPP values, we applied it to a range of smaller models in the Pythia suite. In \Cref{fig:compression-comparison-trends} (left), we show that the log-loss for our method starts out much closer to the original model than AQLM enhanced with PV tuning \citep{malinovskii2024pv}, a state-of-the-art quantization method. Performance of our method smoothly degrades with an elbow around $0.1$ BPP.

\paragraph{Larger models are potentially easier to compress.}
In \Cref{fig:compression-comparison-weights}, we show MSE rate distortion curves for models of various sizes. Interestingly, these reveal that within the Pythia family, the MSE of compressed models gradually improves as model size increases. This is surprising, as MSE is not dependent on the number of parameters. Even more surprising is the striking change in the Pareto frontier between Llama-3-8B and Llama-3-70B, the latter of which maintains very low MSE down to $0.1$ BPP with almost no degradation. These results provide evidence that larger models are easier to compress with our method, implying that the relative benefit of compression during exfiltration attacks may increase as models continue scaling up.

\subsection{Improved LLM Compression}

\paragraph{Method.}
In our method, we employ a compression technique adapted from additive quantization, a generalization of k-means clustering. 
Specifically, we begin by splitting the weights of each layer of the large language model (LLM) into consecutive groups by rows and columns. 
For each group, we apply k-means clustering to compress the weights into vectors derived from a codebook \citep{jegou2010product}.
Following this initial compression, we subtract the reconstructed weight from the original weight and repeat the compression process with the residual for 1 to 4 iterations \citep{egiazarian2024extreme}. After compressing the residuals, we apply a one-dimensional k-means clustering to learn a unique scaling factor for each group, further optimizing the storage requirement.

To determine optimal hyperparameters for compression at various bits per parameter values, we perform a hyperparameter sweep using MSE to the original weights as the metric for selection. We then computed the Pareto frontier and used points on this frontier to select hyperparameters for the main experiments. We visualize several such curves in \Cref{fig:kmeans-comparison} in the appendix.

\begin{table*}[t]
\vspace{-10pt}
\centering
\renewcommand{\arraystretch}{1.2}
\begin{tabular}{@{}lcccc@{}}
\toprule
& \multicolumn{2}{c}{\textbf{Qwen2-1.5B}} & \multicolumn{2}{c}{\textbf{Qwen2-7B}} \\
\cmidrule(lr){2-3} \cmidrule(l){4-5}
\textbf{Method} & \textbf{MMLU} & \textbf{C4 Loss} & \textbf{MMLU} & \textbf{C4 Loss} \\
\midrule
Random Weights & 25.6 & 6.07 & 27.4 & 6.0 \\
Original Model (no compression) & 56.5 & 3.28 & 70.5 & 2.55 \\
\midrule
\multicolumn{5}{l}{\textit{Ours (VQ then train)}} \\
Low compression & 48.4 & 4.27 & 48.5 & 3.22 \\
Medium compression & 54.9 & 3.69 & 61.2 & 3.16 \\
High compression & 55.5 & 3.33 & 65.5 & 2.81 \\
\midrule
\multicolumn{5}{l}{\textit{AQLM}} \\
Low compression & 24.1 & 8.49 & 34.0 & 3.37 \\
Medium compression & 23.5 & 6.39 & 60.3 & 3.28 \\
High compression & 50.7 & 3.65 & 65.6 & 2.74 \\
\bottomrule
\end{tabular}
\caption{Comparison of language model compression methods for Qwen2-1.5B (with a budget of 1 billion tokens, and compression levels of 0.58, 1.15, and 2.03 BPP) and Qwen2-7B (with a fixed budget of 50 million tokens and compression levels of 1.5, 2.26, and 3.01 BPP). Our VQ-then-train method achieves higher performance than AQLM at equivalent compression levels, particularly at lower bits-per-parameter.}
\label{tab:compression-comparison-combined}
\vspace{-10pt}
\end{table*}

Decompression involves reconstructing the full parameter vector from its quantized representation. To ensure the functionality of the decompressed model, we perform fine-tuning initially on a pre-training dataset, followed by instruction tuning on a specific dataset if the original model had been instruction-tuned. Unlike previous quantization schemes that require the fine-tuned model to remain quantizable to the same bit depth, our method does not impose such restrictions, allowing for more flexibility in restoring the model's performance.

\paragraph{Setup.}
We compress Qwen2-1.5B \citep{yang2024qwen2} at three different bits per parameter (BPP) levels, and we use a fixed token budget for decompression. As a baseline, we compare to \textit{Original Model} with no compression and \textit{AQLM} \citep{egiazarian2024extreme}, a state-of-the-art LLM quantization method for efficient inference. For reference, we also show the performance of randomly initialized weights after applying the same fine-tuning procedure used in our method (\textit{Random Weights}). We evaluate baselines and our method using three metrics: bits per parameter (BPP), MMLU accuracy, and C4 log loss.

Additionally, we conduct compression experiments on Qwen2-7B model \citep{yang2024qwen2}\, compressing to 1.5 BPP, 2.26 BPP, and 3.01 BPP. 
The primary distinction between this setup and the previous experiments is the use of a significantly smaller fine-tuning dataset, limited to 50 million tokens. 
Similarly to the Qwen2-1.5B model, we also evaluated the baselines using the same metrics (BPP, MMLU accuracy, and C4 log loss).

For decompression of Qwen2-1.5B, we perform continued pre-training and supervised fine-tuning on RedPajama \citep{NEURIPS2024_d3449733} and Magpie \citep{xu2024magpie}, respectively. We train for a combined $1$ billion tokens on these two datasets, comprising approximately $0.01\%$ of the Qwen2 pre-training data of $7$ trillion tokens.

\begin{figure*}[t]
    \centering
     \includegraphics[width=0.75\textwidth]{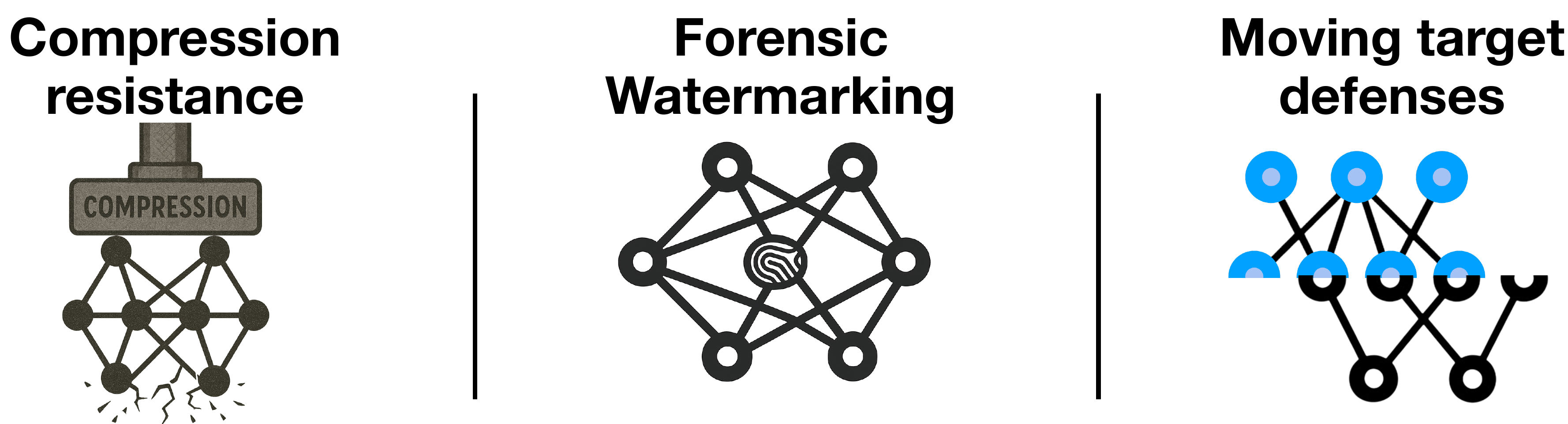}
    \caption{We study three defenses against weight exfiltration: \textbf{Left:} fine-tuning weights to make them harder to compress-to-steal; \textbf{Middle:} adding watermarks to weights to attribute the data and time of exfiltration; \textbf{Right:} a moving target defense that attempts to use model symmetries to make the correct weights more difficult to piece together.}
    \label{fig:weight-defenses}
\end{figure*}

\paragraph{Results.}
Our results for Qwen2-1.5B are in \Cref{tab:compression-comparison-combined}. With AQLM, model performance sharply drops off below $2.03$ bits per parameter (BPP), while our method has much more graceful degradation. Our method maintains reasonable fidelity at $1.15$ BPP, with only a $1.6\%$ reduction in accuracy on MMLU. Thus, our method is able to obtain $16\times$ compression compared to the original model without reducing downstream performance. The Random Weights baseline obtains random chance performance, indicating that the amount of fine-tuning performed during decompression does not introduce new knowledge, and using the compressed weights as a starting point is crucial.

For the larger Qwen2-7B model, our method achieves an MMLU accuracy of 48.5 at 1.52 bits per parameter (BPP), compared to 34.0 for AQLM, representing a difference of 14.5 points. Additionally, our method generally achieves lower C4 log loss values than AQLM across multiple compression levels. Specifically, at 1.52 BPP, our method achieves a C4 log loss of 3.22 nats, whereas AQLM achieves 3.37 nats, indicating a modest improvement in language modeling performance. 

These results demonstrate that when constraints on decompression time are relaxed, models can be compressed to far smaller sizes than was previously known. In \Cref{app:rate_distortion}, we show additional results that suggest larger models may be even easier to compress using our method. In particular, these results suggest that it may be possible to compress Llama-3-70B down to $0.1$ BPP using our method, a $160\times$ compression. This is highly concerning for the setting of exfiltration attacks.


\subsection{Effect of Compression on Weight Exfiltration}
\label{sec:quantitative_model_results}

In \Cref{fig:exfil-dumbbell}, we show the effect of model compression on time until exfiltration and probability of attack success across various scenarios. Each scenario varies the number of compromised daily users, the steganography ratio, and the usage profile. We explore reasonable ranges for these values, discussed in \Cref{sec:quantitative_model}. We find that model compression significantly affects the feasibility of an attack, reducing exfiltration times by multiple orders of magnitude and increasing the probability of attack success.

In \Cref{fig:model-exfiltration}, we directly visualize the amount of bits extracted as a function of time under our quantitative model. This clarifies that besides model compression, other factors, like the number of user accounts and output modality of the model(s), can have a large impact on exfiltration risk. Addressing these other factors could be a feasible direction for improving exfiltration defenses. We now turn to discussing defenses against LLM weight exfiltration.

\vspace{-3pt}
\section{Defenses}
\vspace{-5pt}
\label{sec:defenses}

We consider three defenses that harden against three separate attack vectors: i) making models harder to compress, ii) making them harder to `find,' and iii) making them easier to attribute post-attack.
These are shown schematically in \Cref{fig:weight-defenses}.

\subsection{Finetuning compressibility resistance}\label{resistance}
We identify model compressibility as an important factor in weight security. 
To mitigate model compressibility, we apply finetuning for \textit{compressibility resistance}. 
In particular, we finetune Qwen2.5-1.5B with an off-diagonal covariance penalty added to the standard task loss.  
This makes the weight elements less correlated, and therefore harder to compress.
For each weight matrix \(W\in\mathbb{R}^{d_o\times d_i}\), let
\[
\mu = \frac{1}{d_i}W\,\mathbf{1},\qquad
\Sigma = \frac{(W - \mu\mathbf{1}^\top)\,(W - \mu\mathbf{1}^\top)^\top}{d_i - 1}
\]
be the row-wise covariance of \(W\).  We then include the regularizer $\mathcal{R}(W) \;=\; \sum_{i\neq j} \Sigma_{ij}^2$ in the training objective, which explicitly penalizes pairwise correlations between rows of \(W\).  
The intuition here is that less correlated rows look more random and are thus harder to compress.
Finetuning for 100M tokens with this penalty yields a modest \(~3-10\%\) increase in compressibility cost.

\begin{figure}
    \centering
    \includegraphics[width=\linewidth]{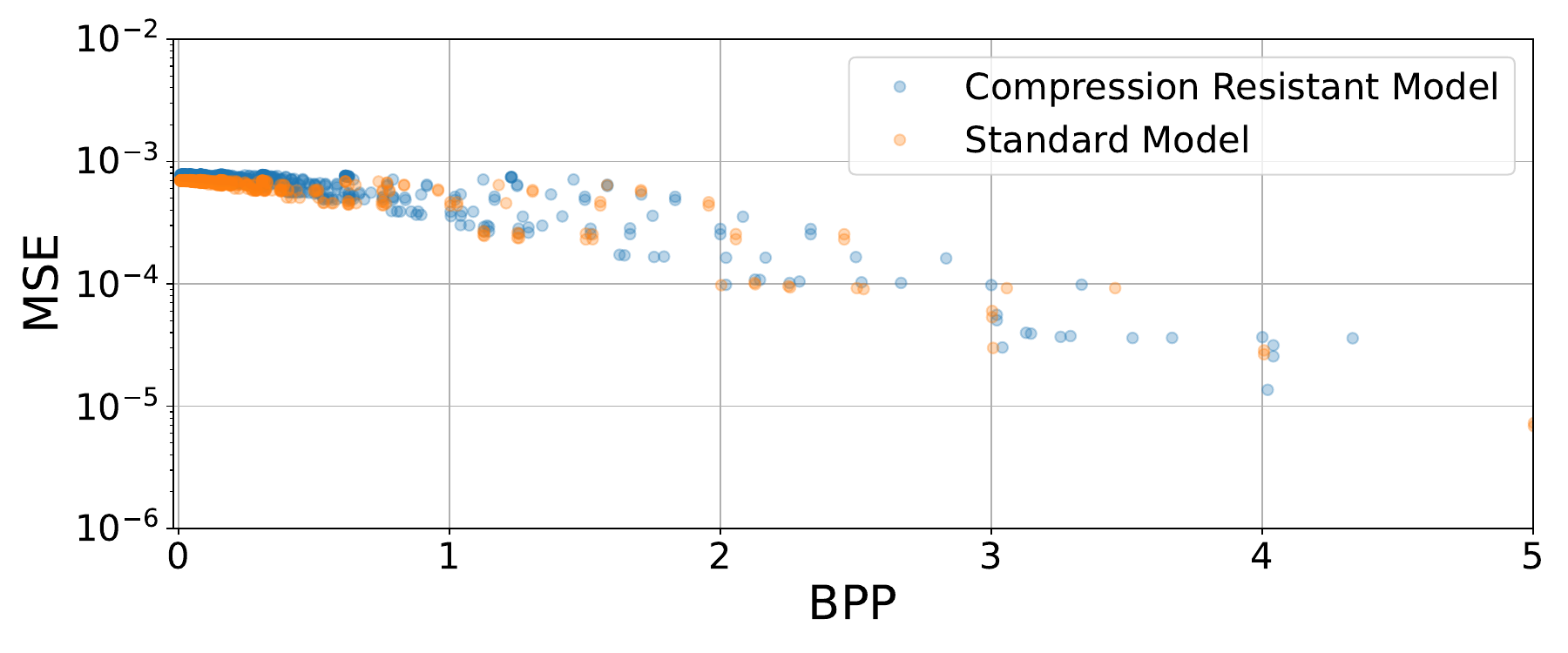}
    \caption{Full MSE vs BPP (rate-distortion) plot of a standard Qwen2.5-1.5B vs a Qwen2.5-1.5B model layers finetuned to be harder to compress.}
    \label{fig:mse-vs-bpp-resistant}
\end{figure}

\begin{figure}
    \centering
    \includegraphics[width=\linewidth]{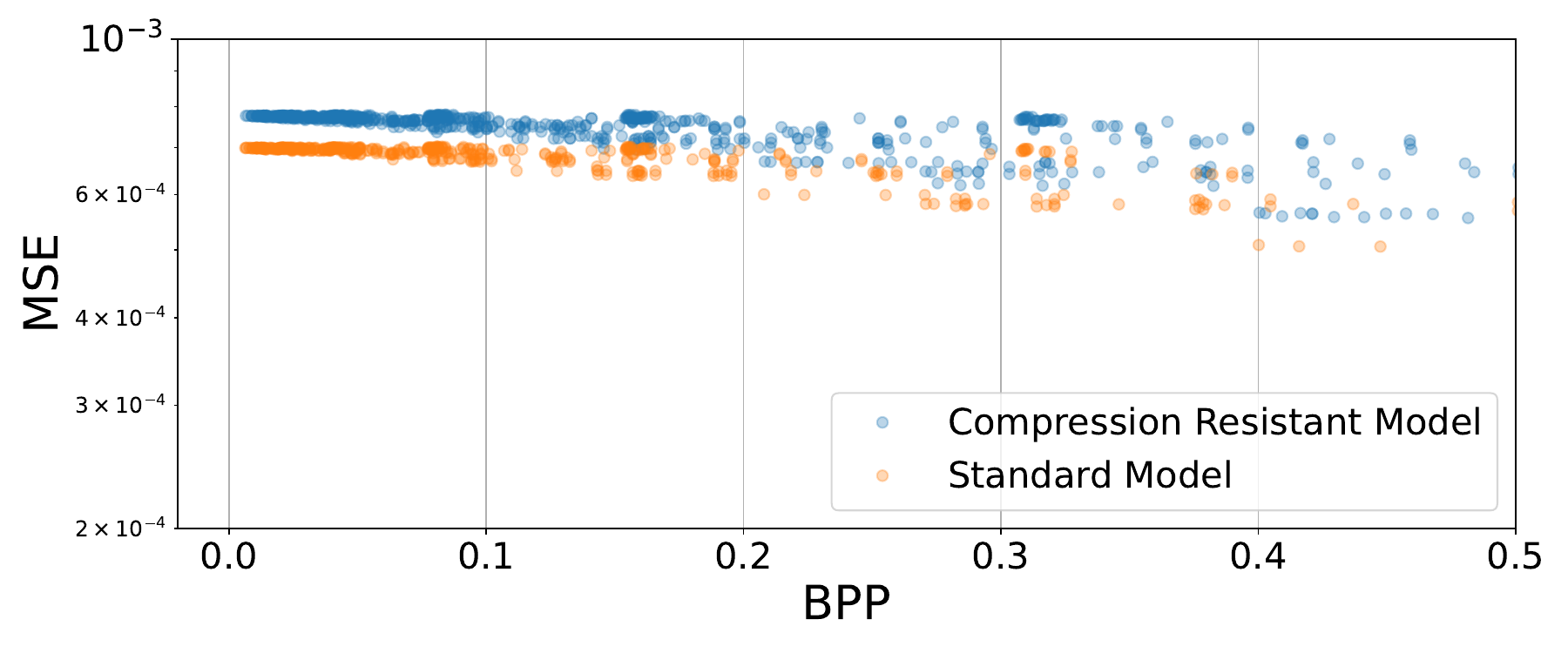}
    \caption{At smaller bits per parameter, the compression resistant model is nearly always more difficult to compress (the same data as \Cref{fig:mse-vs-bpp-resistant} above). MSE vs BPP (rate-distortion) plot of the standard Qwen2.5-1.5B vs a Qwen2.5-1.5B model finetuned to be harder to compress.}
    \label{fig:mse-vs-bpp-resistant-zoom}
\end{figure}

As noted, we finetune a Qwen2.5-1.5B model using an objective designed to make the weights harder to compress by making the neurons less correlated while maintaining the same training loss.
We plot the bits-per-parameter vs mean squared error between the standard and finetuned (`Compression Resistant') model in \Cref{fig:mse-vs-bpp-resistant}.
As noted, we find a modest (but noisy) increase in how difficult the new model is to compress across all compression levels.
The largest and most consistent gains are at lower bits per parameter, shown in \Cref{fig:mse-vs-bpp-resistant-zoom}.

\subsection{Moving-target weight transformation}
In cybersecurity, \emph{moving target defenses} continually re-configure a system’s exposed surface to make reconnaissance more difficult for attackers \citep{Jajodia2011,MDPIIntelligentMTD2023}.  
We explore an idea from \citep{shlegeris2023gauge} that protects weights by applying cheap weight-space symmetries that leave the network’s function unchanged, but make it hard to relate two weights between each other.
In particular, the defender's strategy here is to change the weights of the neural network so that, while network \textit{performance} remains exactly the same, the attacker will struggle to `relate' the weights within a layer at successive time steps.
We use continuous rotations of self-attention projections \citep{Entezari2021,Park2024}.  
In the defense, an inference server hosts a neural network instance that is varied at periodic intervals according to the transformation. 
Any parameter fragments leaked in one interval will not align with those from another (due to the transformation), ideally making it so that the attacker cannot `stitch' the weights together in the correct order.

\paragraph{Adaptive attack.}
Unfortunately, an adversary can subvert this defense by running an explicit \emph{canonicalisation} procedure.  
Canonicalisation transforms the weights so that they are in a consistent order, despite the moving target transformation. 
For the rotation example above, the attacker can use the singular value decomposition, which is invariant to transformations.
That is, let $W_Q, W_K \in \mathbb{R}^{d\times d}$ be the canonical query- and key-projection matrices, and let $O \in \mathbb{R}^{d\times d}$ be any invertible matrix.\footnote{It is not necessary, but to preserve weight norms, $O$ might also be constrained to be orthogonal}
A moving target defense can apply the gauge transform
\[
W'_Q = O^{-\top} W_Q, \qquad W'_K = O W_K,
\]
where $O^{-\top}$ denotes $(O^{-1})^\top$. This preserves the attention scores, since
\[
\begin{aligned}
W'_Q{}^\top W'_K
&= (O^{-\top} W_Q)^\top (O W_K) \\
&= W_Q^\top O^{-1} O W_K \\
&= W_Q^\top W_K.
\end{aligned}
\]

Thus the forward map of the attention layer is unchanged, and the model's function remains the same even though the individual weights have moved.

\begin{figure*}[t]
    \centering
    \includegraphics[width=0.8\textwidth]{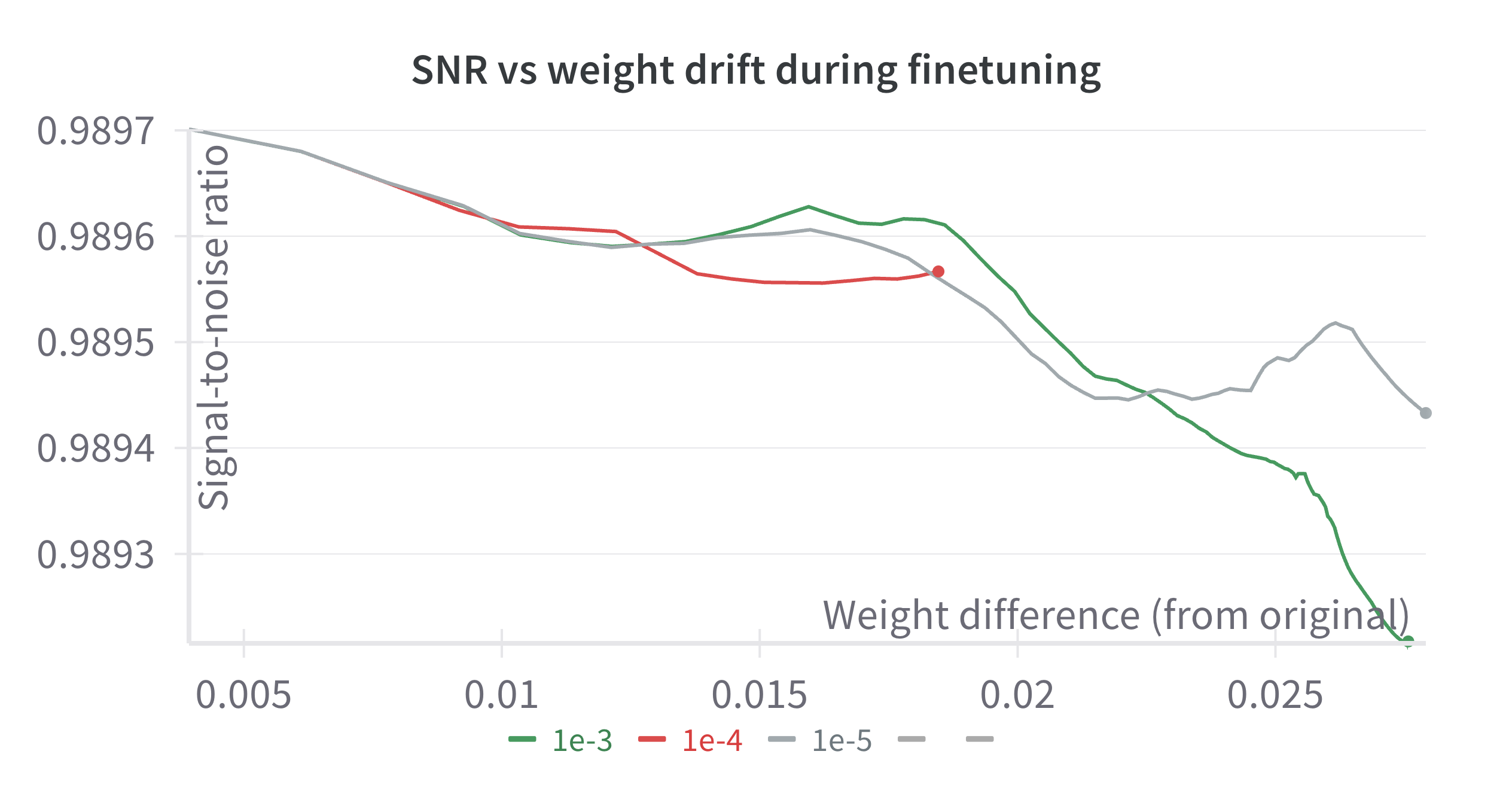}
    \caption{Using existing techniques, defenders can very cheaply add forensic watermarks that are robust to finetuning. Here, we show weight drift (x-axis) vs the signal-to-noise (SNR) ratio (y-axis) during supervised fine-tuning. SNR moves very little over the course of the finetuning run.}
    \label{fig:signal-to-noise}
\end{figure*}

\paragraph{Canonicalisation.}  
This scheme was first proposed in \cite{shlegeris2023gauge}. The attacker first computes the singular value decomposition
\[
W'_Q{}^\top W'_K = U \Sigma V^\top,
\]
then transforms into a canonical basis by setting
\[
W_Q^\ast = W'_Q U, \qquad W_K^\ast = W'_K V.
\]
This yields
\[
W_Q^{\ast\top} W_K^\ast
= U^\top W'_Q{}^\top W'_K V
= U^\top (U \Sigma V^\top) V
= \Sigma.
\]

Since $W_{Q}^{\prime\top} W_{K}^{\prime} = W_Q^\top W_K$, its SVD $(U,\Sigma,V)$ is stable across snapshots. This aligns the singular-vector bases, but not the canonicalised weights $W_Q^* = W_Q' U$ and $W_K^* = W_K' V$, which still depend on $O_t$. Below, we resolve the remaining degrees of freedom.


\paragraph{Recovering useful weights after canonicalisation} However, the attacker still needs to transform \(W_Q'\) and \(W_K'\) into `working' neural network weights. 
We describe one such procedure, inspired by a neural network interpretability method \citep{bansal2021revisiting} in \Cref{sec:stitch-alg}, and run it for a layer of a model to confirm that it is computationally inexpensive.
In short, we canonicalize a layer from Qwen2.5-1.5B and insert it into a weaker model, Qwen2.5-0.5B (along with transformations to map between the different hidden layer sizes of the models).
In \Cref{fig:rotate}, we find that we can recover performance for Qwen2.5-1.5B by recovering useful weights from this layer with only tens of steps.
We also try to recover the full model without stitching, i.e., we canonicalize each layer and attempt to relearn the model end-to-end; this seems to be a much harder learning problem (\Cref{fig:rotate}).
We provide more detail on these experiments in Appendix E.


\subsection{Forensic Watermarking}
We also explore \emph{forensic watermarking}—embedding an owner-controlled signature directly in a weight matrix so that provenance can be verified, even after fine-tuning.
Watermarking is useful, because post-theft attribution allows organizations to identify stolen models in the wild, identify when and where the theft occurred (to e.g. strengthen safeguards), and pursue legal remedies.

Our approach builds on spread–spectrum watermarking for neural networks \citep{uchida2017embedding, pagnotta2022tattooed0}.
First, we encode a 128-bit payload specifying the exact time and origin of the layer (e.g., the datacenter and/or server location) with a BCH(511,447) code.
We spread the code bits across randomly chosen weights of each layer---for our experiments we choose a single layer, \texttt{model.layers.27.mlp.down\_proj.weight}, under the assumption that other layers will behave similarly. 
After running hyperparameter sweeps, we set a per-weight amplitude of \(\gamma = 6\sigma/\sqrt{R}\), where $\sigma$ is the layer’s weight standard deviation and $R$ the number of selected weights.  
To make the watermark \emph{robust} to subsequent supervised fine-tuning, we compute a single backward pass on the input and output of the layer an a single batch of RedPajama tokens, and project the watermark to be orthogonal to that gradient.
This extends the scheme in \citet{huang2024orthogonal} to the case of single-layer watermarking. On an NVIDIA A100, the watermark can be computed in 0.1s (assuming the weights and data are pre-loaded). In our experiments, the raw bit-error-rate remains below $3\%$ after 1.2M‐token supervised finetuning with the Magpie dataset \citep{xu2024magpie0} at learning-rates $1\times10^{-5} - 1\times10^{-3}$. The BCH decoder therefore recovers the full payload with $0$ errors (see \Cref{fig:signal-to-noise} for the signal to noise throughout training). This suggests that weight attribution can succeed even when the attacker chooses to further finetune the model.
However, we do not test robustness to adversarial removal strategies.

\subsection{System-Level Defenses}
So far, we have discussed several defenses that modify the parameters of the model under threat to mitigate exfiltration risks in various ways. Another line of defense involves standard system-level defenses designed for Advanced Persistent Threats (APTs). While these defenses are highly relevant, weight exfiltration attacks present unique challenges that could benefit from specialized approaches. In \Cref{sec:system_level_defenses}, we discuss several standard defense strategies for mitigating the risks of data exfiltration, along with considerations specific to LLM weight exfiltration.

\section{Conclusion}
In this work, we demonstrated that by relaxing decompression constraints in weight exfiltration scenarios, models can be compressed far more effectively than previously thought. Our results show that such aggressive compression greatly reduces the time required for model exfiltration, making attacks significantly more feasible. Towards mitigating this threat, we consider three defenses, two of which attempt to change structural properties of the weights to make them harder to steal along with a moving target defense. These findings emphasize the need for further investment into securing model weights as AIs become increasingly critical assets for both industry and national security.

\section*{Acknowledgments}
We thank Aaron Scher, Adam Karvonen, Alex Cloud, Gabriel Mukobi, Nathaniel Li, Max Lamparth, Steven Basart, and Refine.ink for helpful discussions and feedback on this work. 
This work was supported by the MATS program, Open Philanthropy, and compute resources from the Center for AI Safety.



\appendix
\cleardoublepage


\begin{figure*}[h]
    \centering
    \begin{subfigure}[b]{0.46\textwidth}
        \centering
        \includegraphics[width=\textwidth]{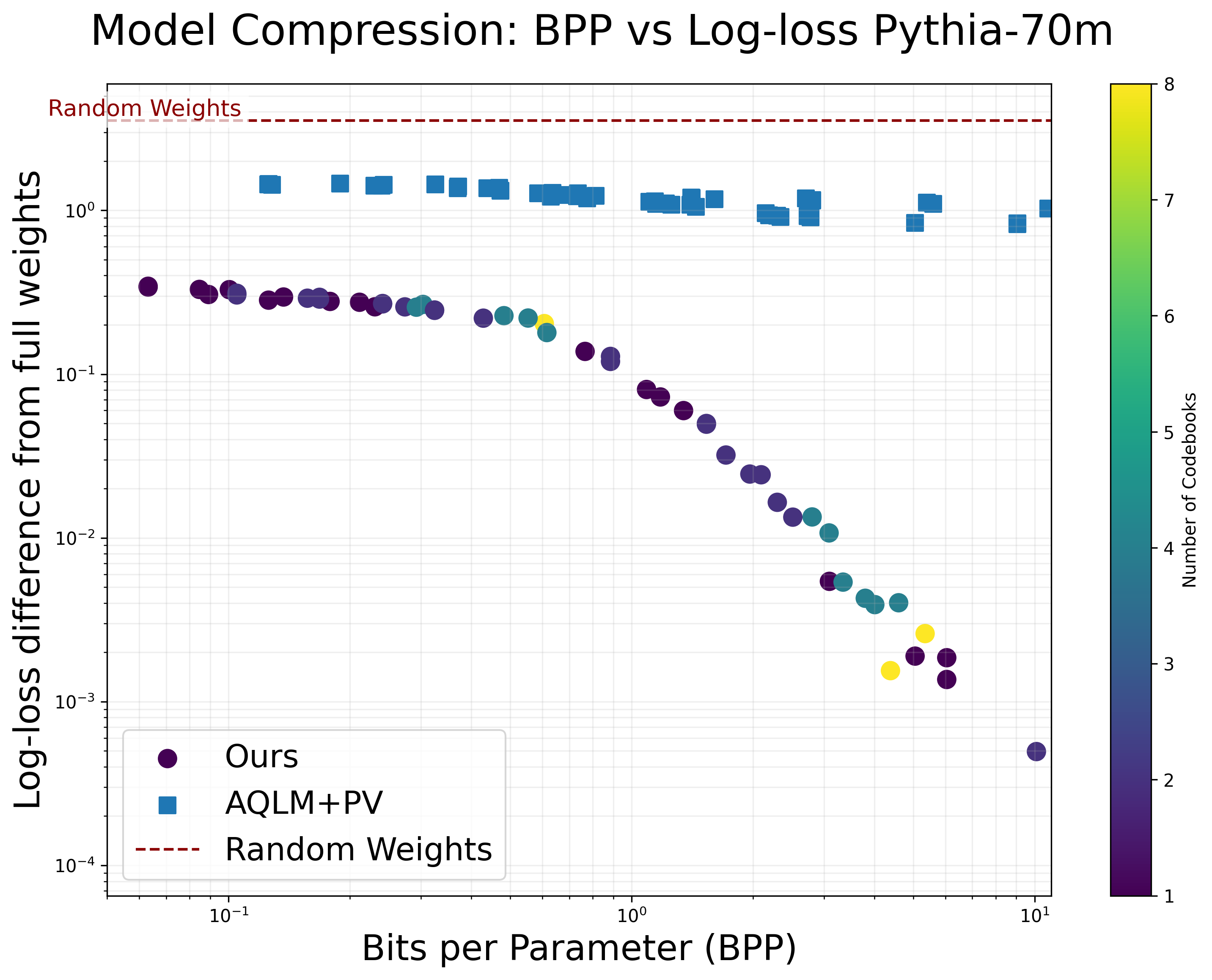}
        \caption{Pythia-70m compressed and then trained with a fixed token budget (8k C4 samples). }
        \label{fig:compression-comparison-70m}
    \end{subfigure}
    \hfill
    \begin{subfigure}[b]{0.5\textwidth}
        \centering
        \includegraphics[width=\textwidth]{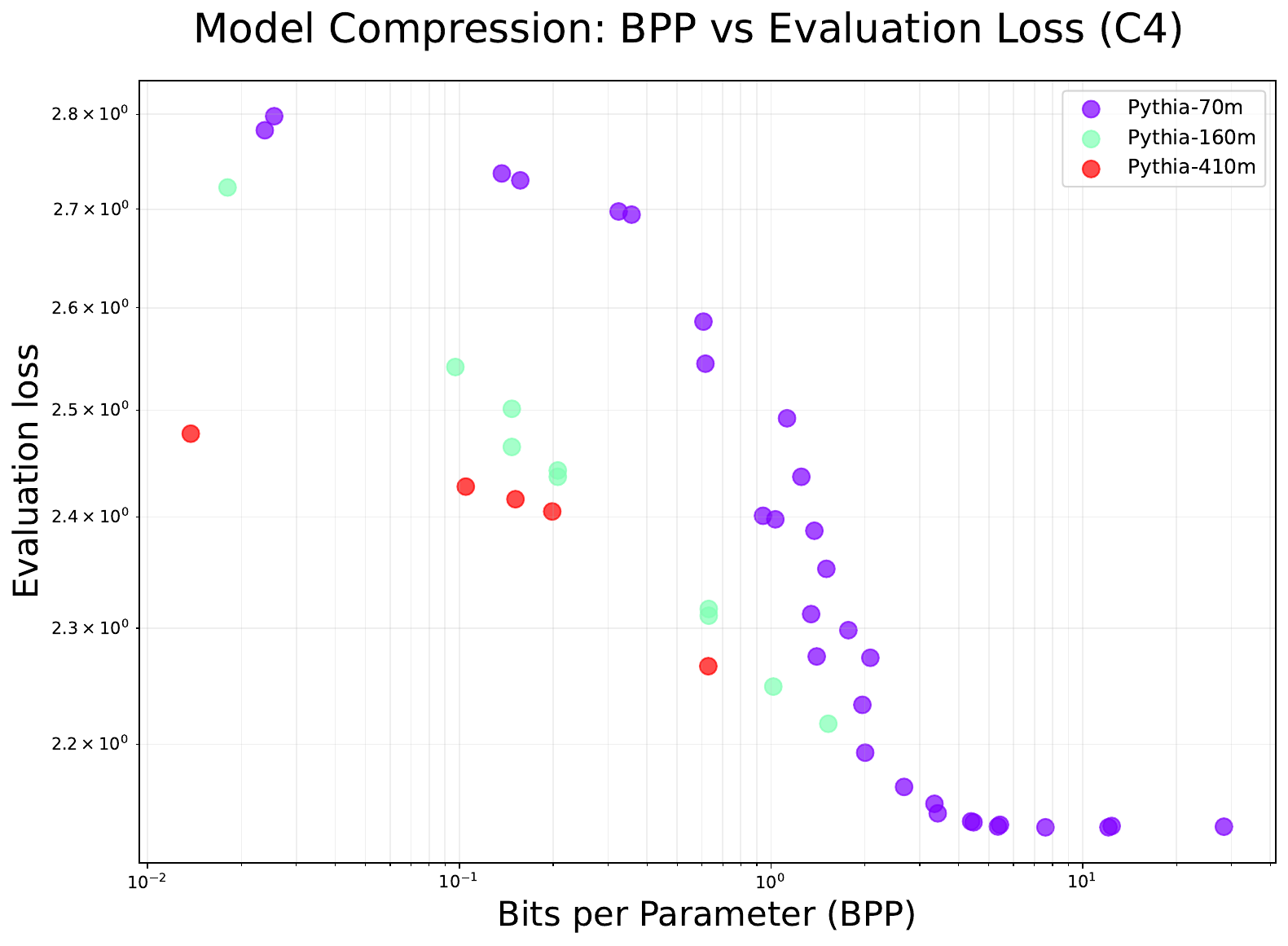}
        \caption{Three models from the Pythia suite compressed and trained with 150K C4 samples.}
        \label{fig:compression-comparison-suite}
    \end{subfigure}
    \caption{Compressing Pythia models to various bits-per-parameter (BPP) and recovering performance with continued training. For Pythia-70m, there is an elbow for both plots at approximately $0.5$-$0.6$ BPP. The trend is less clear for the larger Pythia models. See \Cref{fig:compression-comparison-weights} for BPP vs MSE in weight-space. }
    \label{fig:compression-comparison-trends}
\end{figure*}


\begin{figure*}[t]
    \centering
    \includegraphics[width=\textwidth]{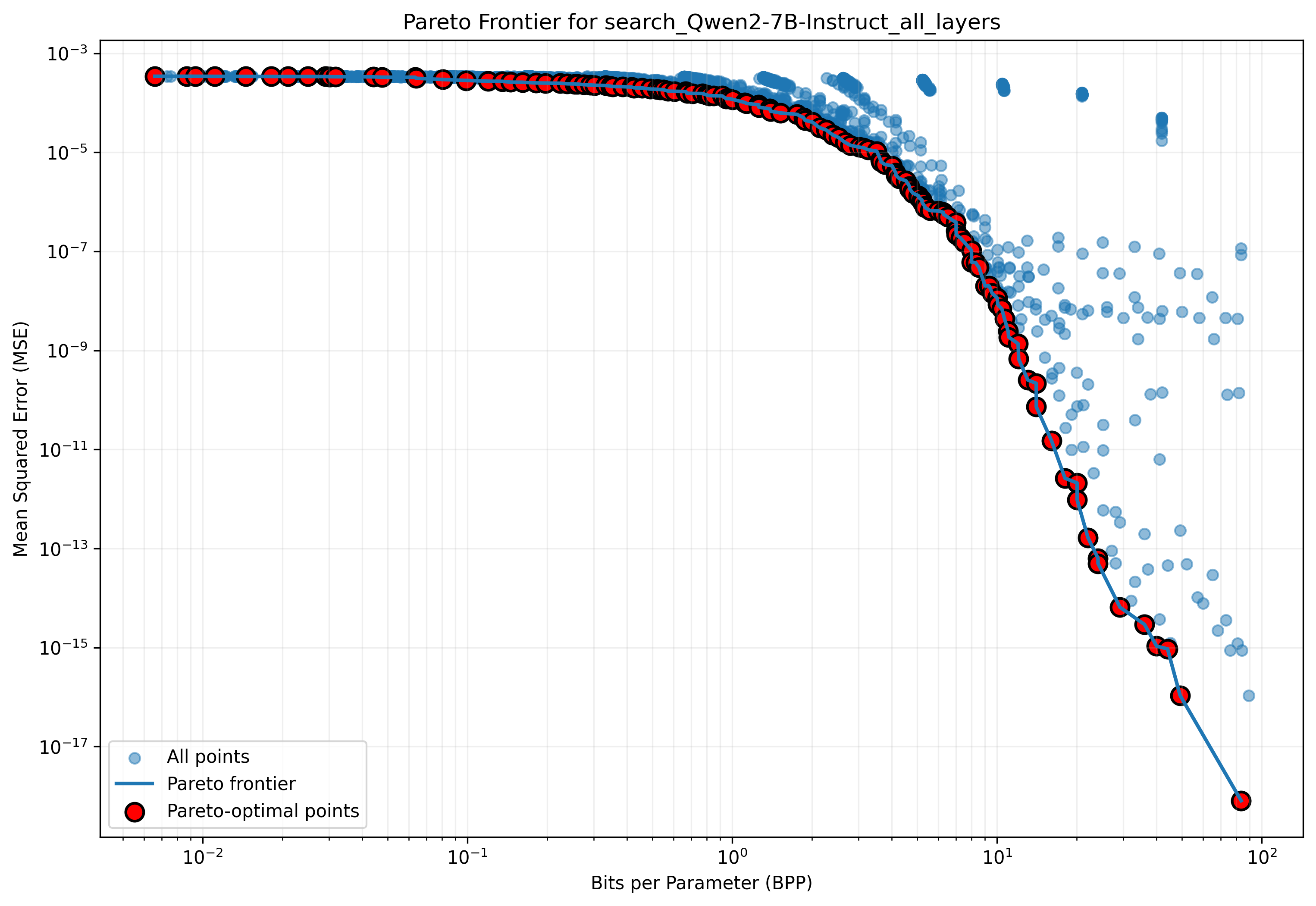}
    \caption{The MSE between the original and compressed weights vs the bits-per-parameter (BPP), along with the Pareto frontier for MSE vs BPP, for Qwen2-7B.}
    \label{fig:compression-comparison-weights-Qwen2-7b}
\end{figure*}

\begin{figure*}[ht!]
    \centering
    \begin{subfigure}[b]{0.49\textwidth}
        \centering
        \includegraphics[width=\textwidth]{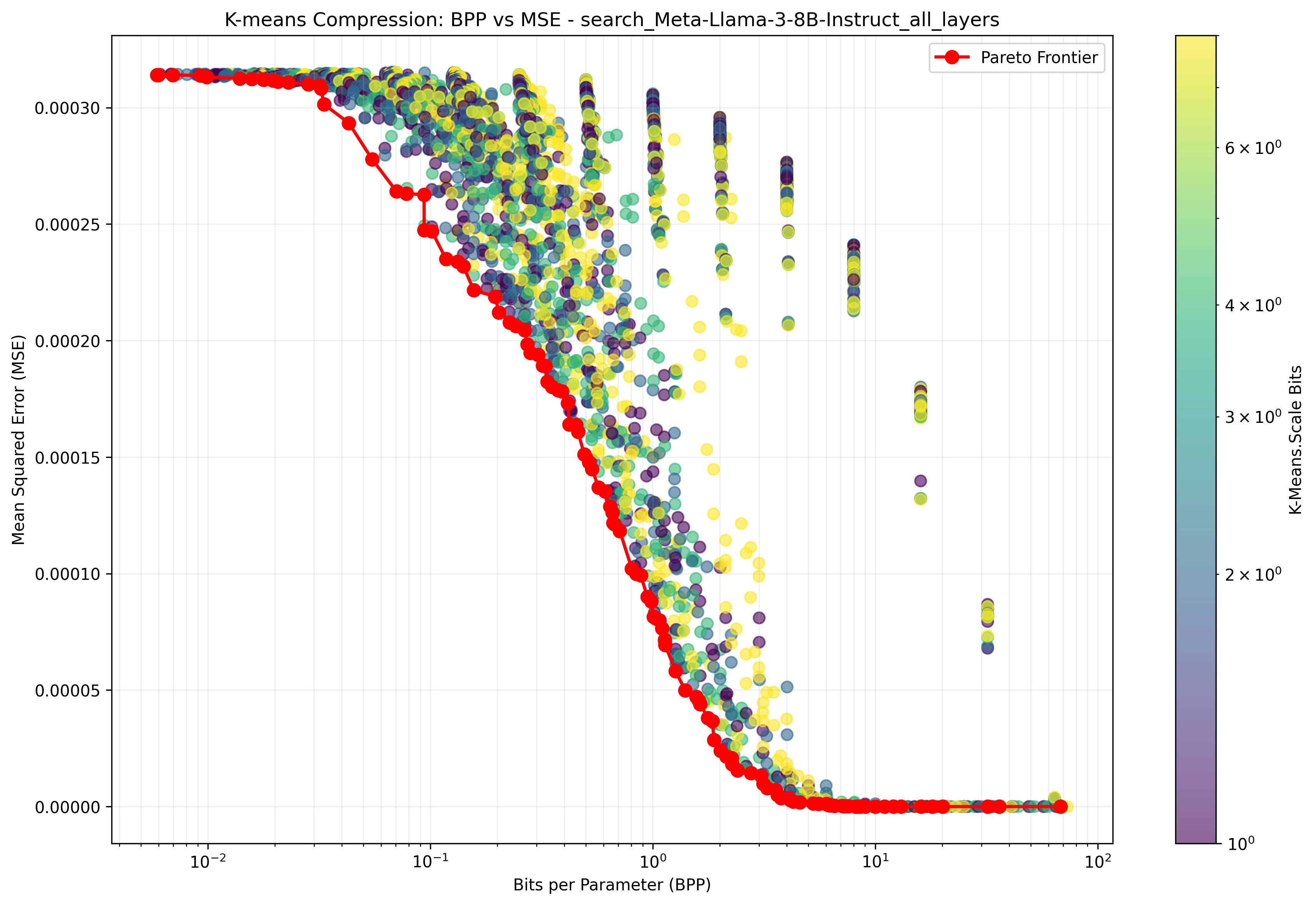}
        \caption{Llama-3-8B}
        \label{fig:kmeans-8b}
    \end{subfigure}%
    \hfill
    \begin{subfigure}[b]{0.49\textwidth}
        \centering
        \includegraphics[width=\textwidth]{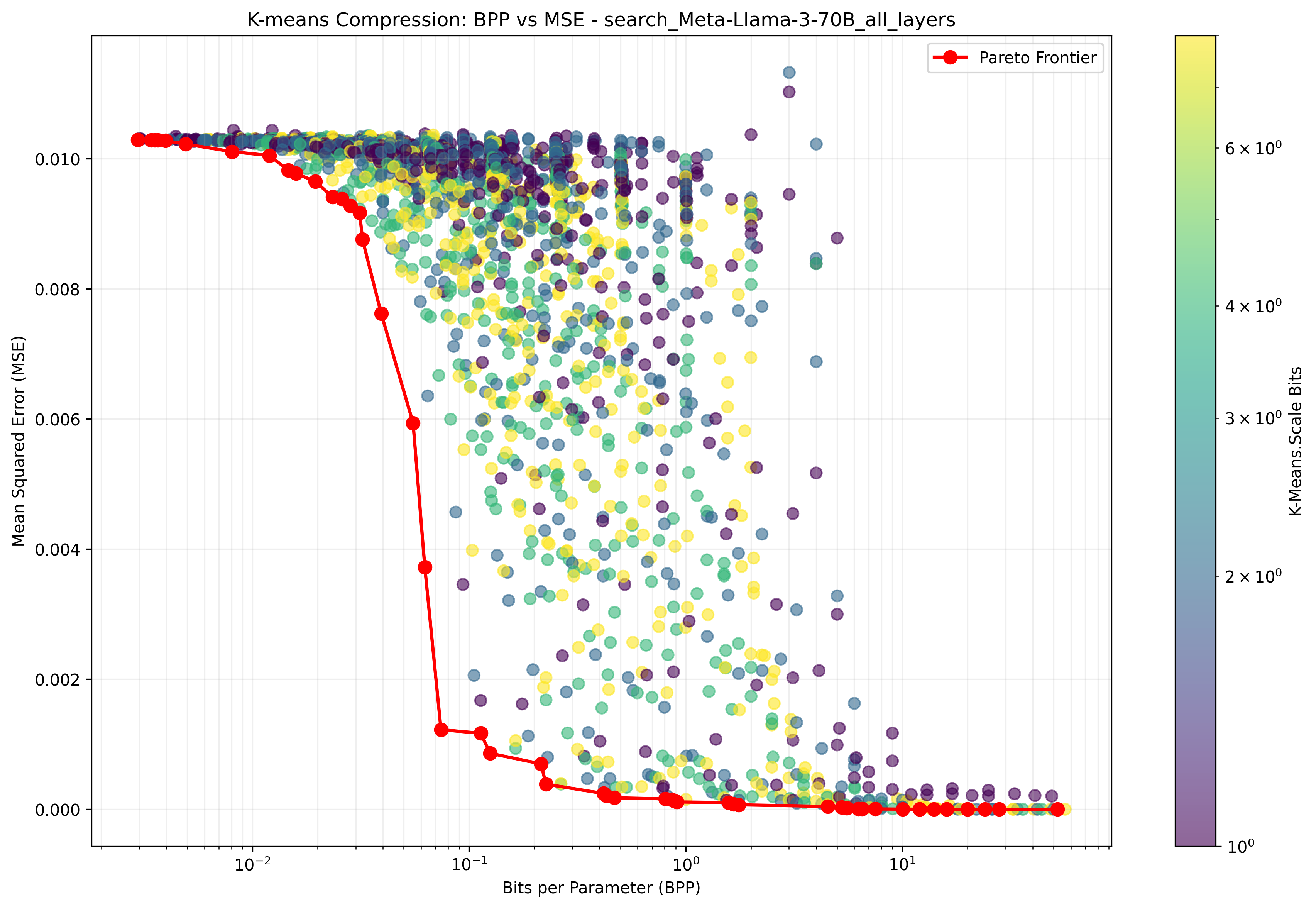}
        \caption{Llama-3-70B}
        \label{fig:kmeans-70b}
    \end{subfigure}%
    \\
    \begin{subfigure}[b]{0.49\textwidth}
        \centering
        \includegraphics[width=\textwidth]{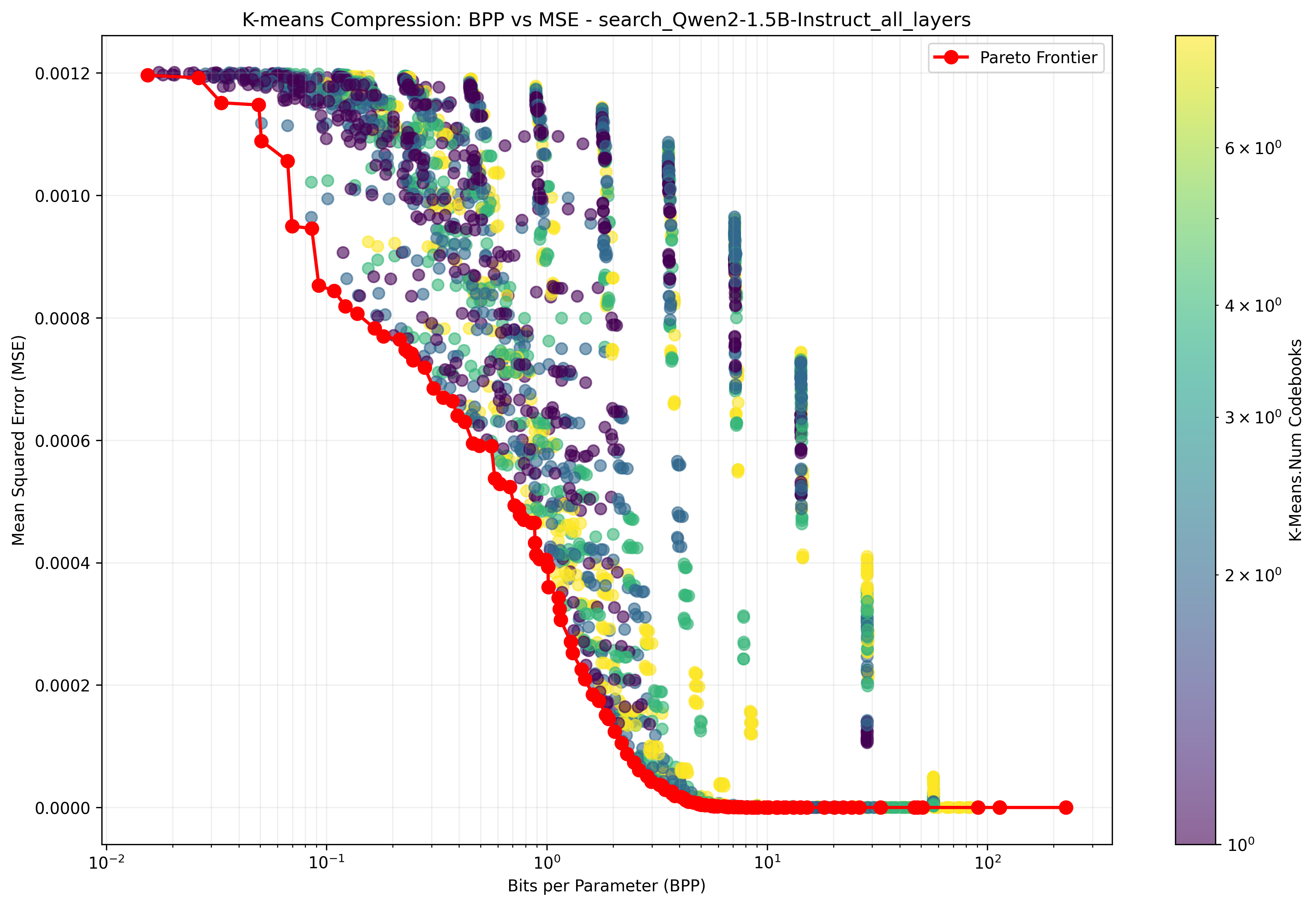}
        \caption{Qwen2-1.5B}
        \label{fig:kmeans-qwen2}
    \end{subfigure}
    \caption{Comparison of MSE rate-distortion curves on different models. Strikingly, the Llama-3-70B Pareto frontier obtains very high-fidelity compression down to $0.1$ bits per parameter (BPP), representing a $160\times$ compression. Note that we do not run full evaluations on Llama 3 models due to resource constraints.}
    \label{fig:kmeans-comparison}
\end{figure*}


\section{Ethical Considerations}
\label{sec:ethical-consdirations}
{This work has dual-use implications. We aim to support defenders by quantifying exfiltration feasibility and studying mitigations including egress limits, reducing compressibility, moving target defenses, and forensic watermarks. }

\section{System-Level Defenses}
\label{sec:system_level_defenses}

Here, we discuss additional defense strategies against exfiltration attacks that do not involve modifying model weights. These defenses are system-level, as opposed to the model-level defenses described in \Cref{sec:defenses}.

\paragraph{Anomaly detection and monitoring.} Traditional APT detection methods typically rely on general anomaly detection, scanning for statistical deviations over large quantities of server activity and network logs. Weight exfiltration, however, occurs in a highly orchestrated environment where server behavior is predictable and narrowly focused. Inference servers handle a single task: processing user inputs and generating outputs in a consistent format. This makes it possible for defenders to apply simple packet inspection techniques to detect anomalies. However, a key challenge lies in the high variance of resource usage in LLM inference, which could allow adversarial activity to be obscured by normal fluctuations in system load.

\paragraph{Hardware-based security measures.} Hardware security features embedded in GPUs and CPUs can play a critical role in defending against weight exfiltration. Some NVIDIA GPUs, for instance, implement secure boot mechanisms that use cryptographic verification to ensure that only authorized firmware and software can be executed on the hardware. Although these capabilities are not universally implemented across all inference servers and may have limitations, they offer a promising layer of defense by preventing unauthorized access to model weights at the hardware level.

\paragraph{Upload rate limits.} One of the most straightforward and effective defenses is to impose strict upload rate limits on the server, ensuring that data egress is constrained to the minimum required for legitimate user workloads. This was recently proposed by \citep{greenblatt2024preventing}, who also suggest compressing model outputs to restrict total data egress. Our quantitative model already accounts for such limits, but if these constraints are absent, exfiltration could become far easier for attackers. 

\paragraph{Internal data transfer limits.} To further mitigate the risk of weight exfiltration, internal data transfer limits between system components could be enforced. In particular, many LLM-based systems include additional modules for tasks such as voice or image generation, which can require higher data transfer rates than text-based models. Our results in Section 4.3 in the main paper suggest that high network requirements for these additional modalities greatly increase exfiltration success rates. However, this assumes that servers where LLM weights are stored would have high internal bandwidth to servers that integrate different modalities and send outputs to users. By physically decoupling the LLM servers from where final integrated outputs are sent to users, exfiltration data rates could be effectively limited to the core LLM's outputs.

\section{Limitations}
\label{app:limitations}
In this section, we discuss the limitations of our threat model and analysis. While we present a preliminary investigation of how model compression increases the risk of weight exfiltration, further research is needed to fully understand and model this threat. Some key limitations are described below.

Our threat model does not account for how increased activity by the attacker might increase the rate of detection. We also do not incorporate distributional estimates for the key factors in our quantitative model, although we consider ranges for certain factors. Finally, our model does not factor in the computational costs for the attacker to perform K-means clustering on the infiltrated servers. K-means requires computing and storing cluster centers, which could be detectable by the defender. However, the attacker might mitigate this by running K-means iteratively, uploading chunks of the weights and only storing a small number of cluster centers at a time. By doing so, they could adjust their resource usage to stay below detection thresholds.

\begin{figure*}[t]
    \centering
    \includegraphics[width=\textwidth]{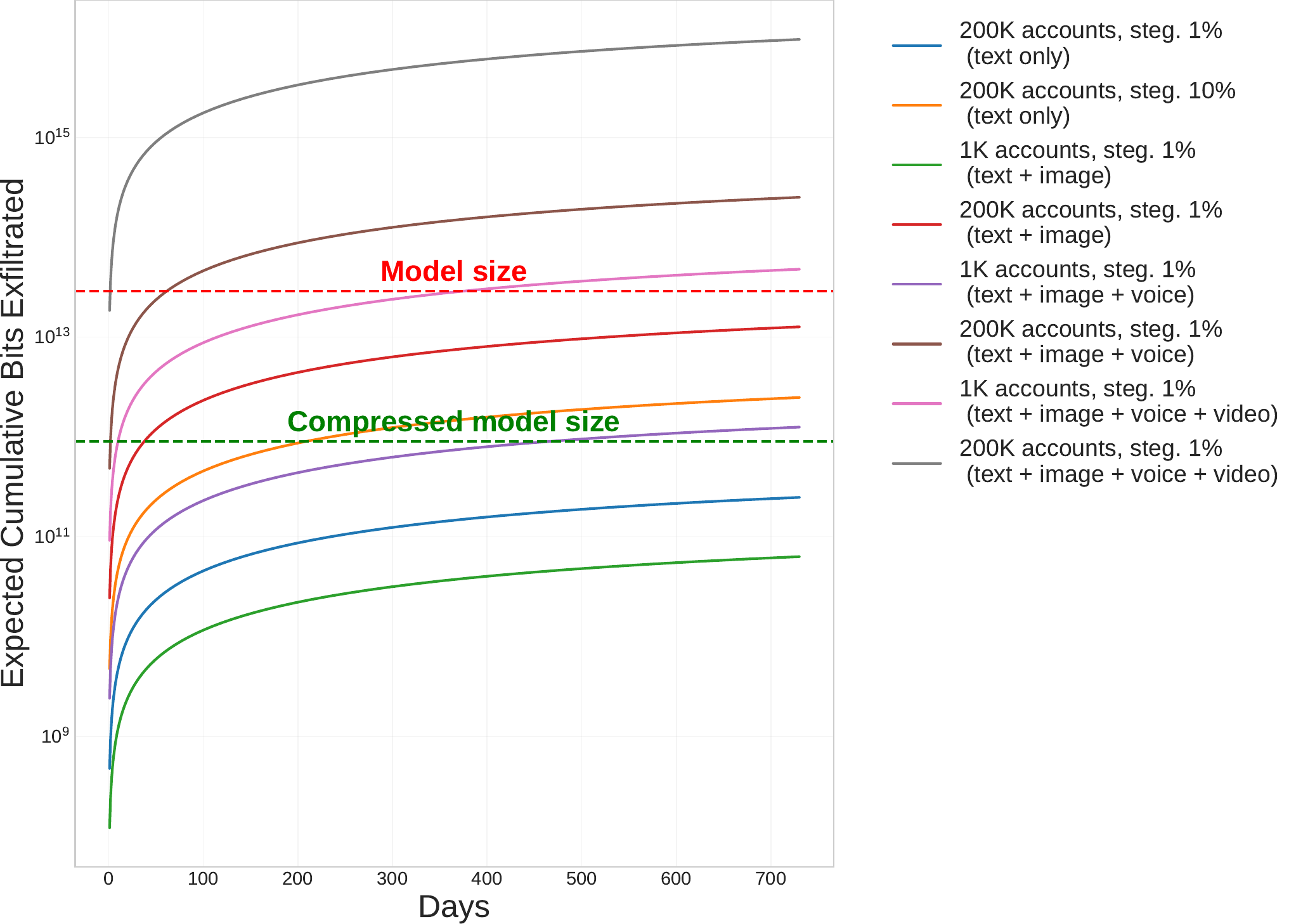}
    \caption{Data exfiltrated over time across different estimates for factors in our quantitative model of exfiltration attacks. The original model size is shown in red, while the model compressed with the assumed compression ratio of $1/32$ is shown in green. While model compression substantially accelerates attacks, other factors have large effects as well, including the usage profile assumed for the average API user. For example, APIs with text, image, voice, and video generation could be susceptible to far quicker exfiltration due to the higher required data rates for serving this content.}
    \label{fig:model-exfiltration}
\end{figure*}

\section{Estimates for Quantitative Model Factors}
\label{app:estimates}

In \Cref{tab:estimates}, we show our estimates for the factors in our quantitative model. We calibrate these factors to the setting of an exfiltration attack on a model similar to GPT-4. Below, we describe how we arrive at these estimates.

\begin{itemize}[leftmargin=*]
    \item \textbf{Network egress per user per day:} We estimate this using the different usage profiles. See below for details.
    \item \textbf{Number of compromised daily users:} We start from an assumption of approximately $100$ million daily active users for the API and assume that no more than $0.2\%$ of these users could belong to a Sybil attack before suspicious activity would be easily detected. This corresponds to $200,000$ users.
    \item \textbf{Steganography ratio:} For text steganography, values of up to $5$ bits per word are achievable \citep{kang2020generative}. An average English word taking up 5 ASCII characters requires $40$ bits, corresponding to a ratio of roughly $0.1$. Steganography may be even more effective for continuous outputs like images, audio, or video. We use a fixed ratio of $0.1$ for simplicity.
    \item \textbf{Model size:} We assume a model with $1.8$ trillion half-precision floating point parameters, corresponding to $2.88 \times 10^{13}$ bits.
    \item \textbf{Compression ratio:} We use a compression ratio of $1/32$. This is an extreme level of compression that stresses our method, but we still obtain nontrivial performance at this level.
    \item \textbf{Probability of detection per day:} We assume a value of $0.1\%$. This value is highly situational, but we assume attackers would adjust their activity to keep it fairly low.
\end{itemize}

We now describe the estimates of network egress per user per day for each usage profile.

\begin{itemize}[leftmargin=*]
    \item \textbf{Text Only:} We assume each user generates approximately $10$ pages of text per day, each taking up $500$ English words. Each word is $5$ ASCII characters on average with a space, corresponding to $6 \times 500 \times 10 \times 8 = 240,000$ bits.
    \item \textbf{Text + Images:} We assume images are generated at a resolution of $1024 \times 1024$ pixels with a $24$ bit color depth, corresponding to $1024 \times 1024 \times 24 = 25,165,824$ bits. We assume the image is compressed to $1.5$ MB after compression, yielding $1.5 \times 1,000,000 \times 8 = 12,000,000$ bits. This gives a total of $240,000 + 12,000,000 = 12,240,000$ bits.
    \item \textbf{Text + Images + Voice:} We assume $30$ minutes of audio per day and that the data is provided as an MP3 with a bitrate of $128$ kilobits per second. This gives $128 \times 60 \text{ seconds}/\text{minute } \times 30 \text{ minutes } = 230,400$ kilobits, or $230,400,000$ bits. This gives a total of $230,400,000 + 12,240,000 = 242,640,000$ bits.
    \item \textbf{Text + Images + Voice + Video:} We assume $30$ minutes of video per day and that the data is provided at $1920 \times 1080$ resolution at $30$ frames per second with an average bitrate of $5$ megabits per second. This gives $5 \times 60 \text{ seconds}/\text{minute } \times 30 \text{ minutes } = 9,000$ megabits, or $9,000,000,000$ bits. This gives a total of $9,000,000,000 + 242,640,000 = 9,242,640,000$ bits.
    
\end{itemize}

\begin{table*}
\centering
\begin{tabular}{@{}lll@{}}
\toprule
\textbf{Parameter} & \textbf{Assumption} & \textbf{Usage Profile} \\
\midrule
Network egress per user per day & $240,000$ bits & Text Only \\
Network egress per user per day & $12,240,000$ bits & Text+Images \\
Network egress per user per day & $242,640,000$ bits & Text+Images+Voice \\
Network egress per user per day & $9,242,640,000$ bits & Text+Images+Voice+Video \\
Number of compromised daily users & $\leq 200,000$ & All usage profiles \\
Steganography ratio & $\leq 0.1$ & All usage profiles \\
Model size & \(2.88 \times 10^{13}\) bits & All usage profiles \\
Compression ratio & $1/32$ & All usage profiles \\
Probability of detection per day & $0.001$ & All usage profiles \\
\bottomrule
\end{tabular}
\caption{Order-of-magnitude estimates for exfiltrating a $1.8$ trillion parameter language model}
\label{tab:estimates}
\end{table*}

\newpage
\clearpage
\section{Formalizing our Threat Model}
\label{sec:defense_game}
\subsection{Attacker stitching algorithm}\label{sec:stitch-alg}
\begin{algorithm}[H]
\caption{Stitch-and-Relearn Canonicalised Layers}
\label{alg:stitch_relearn}
\begin{algorithmic}[1]
\REQUIRE $\mathcal{M}_\text{large}$ — target backbone (e.g.\ Qwen 2.5-1.5B)
\REQUIRE $\mathcal{M}_\text{small}$ — size-compatible helper model (e.g.\ Qwen 2.5-0.5B)
\REQUIRE $\{(W_Q',W_K')_\ell\}$ — canonicalised projections of layer $\ell$
\REQUIRE SGD hyper-parameters $\eta,\;N_{\text{steps}}$
\STATE
\STATE \textbf{function} StitchAndRelearn($\mathcal{M}_\text{large}, \mathcal{M}_\text{small}, (W_Q',W_K'), \eta, N_{\text{steps}}$)
\STATE \quad \textbf{1. Dimension-bridging rotations.}
\STATE \quad Select $R_Q,R_K$ s.t.\ $W_Q'R_Q$ and $R_K^{\!-1}W_K'$ match the input/output
\STATE \quad dims of the corresponding layer in $\mathcal{M}_\text{small}$.
\STATE
\STATE \quad \textbf{1. Insert canonical layer.}
\STATE \quad Replace the original $(W_Q,W_K)$ of layer $\ell$ in $\mathcal{M}_\text{small}$ by
\STATE \quad $(W_Q'R_Q,\;R_K^{\!-1}W_K')$, where  Freeze all other parameters.
\STATE
\STATE \quad \textbf{3. Rotation-only fine-tune.}
\FOR{$t=1$ \textbf{to} $N_{\text{steps}}$}
\STATE \quad\quad Compute loss on a proxy objective (e.g.\ next-token).
\STATE \quad\quad Update $R_Q,R_K$ with SGD: $R \leftarrow R - \eta\,\nabla_R\!\mathcal{L}$
\ENDFOR
\STATE
\STATE \quad \textbf{4. Transfer to full model.}
\STATE \quad Copy the relearned canonical layer (including $R_Q,R_K$) into the
\STATE \quad corresponding layer of $\mathcal{M}_\text{large}$. Optionally unfreeze
\STATE \quad neighbouring layers and run a brief global fine-tune.
\STATE
\STATE \quad \textbf{return} updated $\mathcal{M}_\text{large}$
\STATE \textbf{end function}
\end{algorithmic}
\end{algorithm}

\begin{figure}[htbp]
    \centering
    \includegraphics[width=0.45\textwidth]{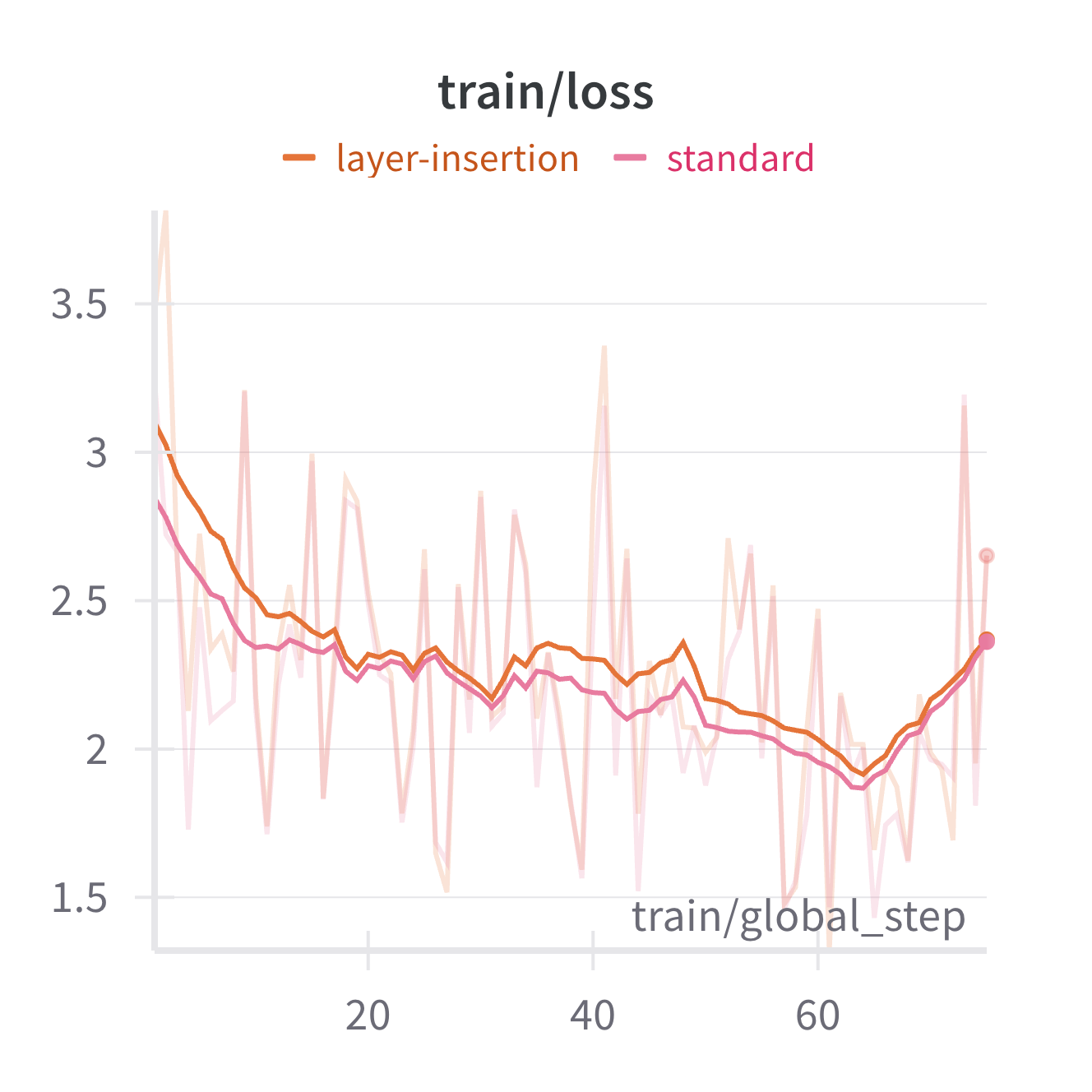}    \includegraphics[width=0.45\textwidth]{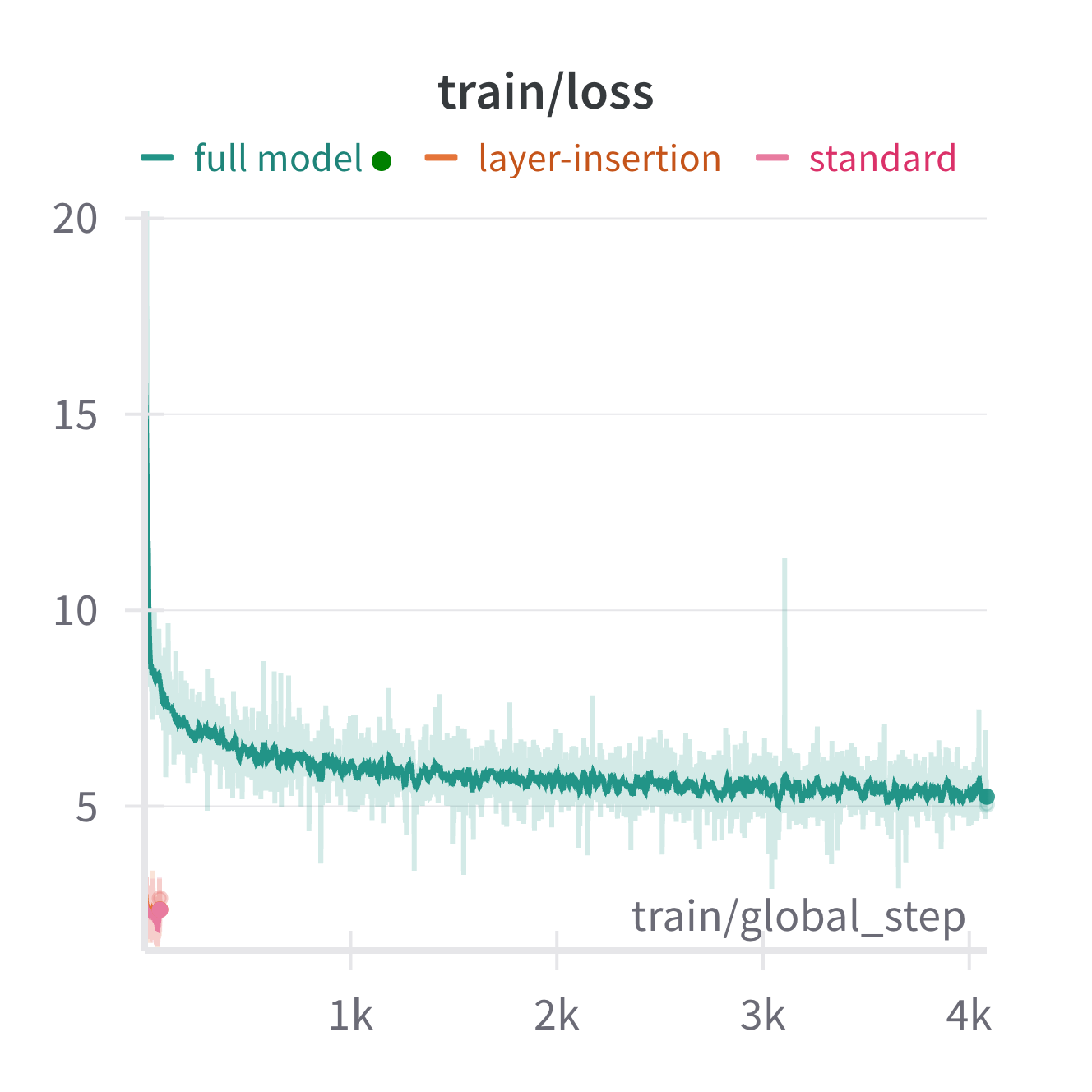}

    \caption{\textbf{Above:} Single layer rotation-only finetunes are cheap. We implement part of the attack steps (1-3 in \cref{sec:stitch-alg}), and find that the aligned layer can be learned quickly-- MMLU performance is also recovered. \textbf{Below:} It is much harder to learn useful weights for the full model at the same time-- this motivates an algorithm like stitching.}
    \label{fig:rotate}
\end{figure}

\subsection{Motivation and attacker model}
We study an adversary who compromises an inference host, must remain under a flagging classifier’s outbound-traffic threshold \(\mathrm{TH}\), and operates within a dwell-time budget \(T\).  
Their objective is to exfiltrate a representation \(\mathcal{E}\) that, after at most \(\Phi_{\mathrm{ft}}\) additional FLOPs of fine-tuning, yields a model \(\widehat{W}\) whose downstream performance exceeds a target threshold \(\tau\).  
In the absence of active counter-measures, extreme compression (e.g.\ vector quantisation to \(<\!1\,\mathrm{bit/param}\)) lets the attacker keep \(\lvert\mathcal{E}\rvert\!\ll\!\lvert W\rvert\) and evade detection.

\subsection{Introducing gauge refreshes}
The defender installs a scheduler that, every \(\Delta\) minutes, samples a gauge transform \(g_t\) from a low-cost symmetry, namely rotations of \((W_Q,W_K)\) and \((W_O,W_V)\) of the attention layer, wipes the GPU memory, and pushes the fresh weights \(W_t=g_t(W)\) to the GPU server.  
\paragraph{Original attack.}
During round \(t\):
\begin{enumerate}[leftmargin=*,itemsep=0pt]
  \item Read \(W_t\) and run in-place compression \(\mathcal{C}_t(\cdot)\) to get a code \(\mathcal{E}_t\).
  \item Exfiltrate within the per-round bandwidth cap imposed by \(\mathrm{TH}\).
\end{enumerate}
After \(T' = \lceil T/\Delta\rceil\) rounds the attacker holds fragments \(\{\mathcal{E}_t\}\) plus any alignment metadata.  
The attacker now holds \(T'\) `jumbled' fragments. We claim that as \(T'\) approaches the number of neurons in a model, the attacker incurs the full cost of pretraining the model.

\paragraph{Canonicalisation subverts the moving target defense.}
Unfortunately, described next, \textit{canonicalization} subverts the defense.
Thus the moving-target scheme \emph{alone} is insufficient; additional hardening, for example by injecting small structured noise, may be required.  

\paragraph{Attacker’s new workflow.}
During round \(t\):
\begin{enumerate}[leftmargin=*,itemsep=0pt]
  \item Read \(W_t\) and run in-place compression \(\mathcal{C}_t(\cdot)\) to get a code \(\mathcal{E}_t\).
  \item Perform \emph{canonicalisation}—e.g.\ via SVD or QR decomposition \(\Phi_{\text{align},t}\) FLOPs—to map \(\mathcal{E}_t\) into a basis invariant to gauge refreshes.
  \item Exfiltrate, where it is now not necessary to steal all of the layer weights in time \(\Delta\).
\end{enumerate}
After \(T' = \lceil T/\Delta\rceil\) rounds the attacker holds fragments \(\{\mathcal{E}_t\}\) plus any alignment metadata.  
They merge and fine-tune at cost \(\Phi_{\mathrm{ft}}\) to produce \(\widehat{W}\).

\subsection{Cost-based success condition}
Let
\[
\Phi_{\text{tot}}
=\sum_{t=1}^{T'} \bigl(\Phi_{\mathrm{comp},t} + \Phi_{\text{align},t}\bigr)
      + \Phi_{\mathrm{ft}} .
\]
The attacker \emph{succeeds} if they:
(i) stay below \(\mathrm{TH}\) in every round;  
(ii) keep total wall-clock time \(\le T\); and  
(iii) obtain \(R(\widehat{W})\ge\tau\) with \(\Phi_{\text{tot}} < \gamma\Phi_{\textsc{train}}\) for some advantage factor \(\gamma<1\).

\paragraph{Next steps}
Future work might search for transformations whose inversion provably lacks efficient algorithms. For example, we experimented with strategically adding heavy-tailed and/or sparse noise, and found that it made SVD and QR decompositions increased the reconstruction error of each by nearly 50\%, with only a 6\% cost in MMLU. 
However, a sophisticated attacker could straightforwardly filter for this kind of naive noise.

\section{Watermarking individual layers for attribution}
\label{app:watermark}

For the encoding procedure, let $\mathbf{W}\!\in\!\mathbb{R}^{d}$ be the flattened target weight tensor and
$R\!\ll\!d$ the watermark budget.  We sample a fixed index set
$\mathcal{I} = \{i_1,\dots,i_R\} \subseteq \{1,\dots,d\}$ and a spreading matrix
$\mathbf{S}\!\in\!\{\!-\!1, +1\}^{L\times R}$ with i.i.d.\ Rademacher entries.
A 128-bit Unix timestamp $\mathbf{m}\!\in\!\{0,1\}^{128}$ is zero-padded to 447 bits (message padding)
and encoded with a $\mathrm{BCH}(511,447)$ code, yielding a 511-bit codeword.
This codeword is then zero-padded to $L=640$ bits (length padding) to obtain
$\mathbf{c}\!\in\!\{0,1\}^{L}$. We map $\mathbf{c}\!\mapsto\!\mathbf{b}=2\mathbf{c}-1\in\{\!-\!1,+1\}^{L}$ and set

\[
\Delta w_{i_r}
  = \gamma\sum_{k=1}^{L} b_k S_{kr},\quad
  r = 1,\dots,R, \qquad
  \gamma = \eta\,\sigma_{\mathbf{W}}/\sqrt{R},
\]
with $\eta\!=\!6$ fixed.  All other weights remain untouched.

We update \citet{huang2024orthogonal} to watermark a single layer (instead of the whole network) by considering only the input/output map of the layer of interest.
This also allows for an efficient implementation: for a layer
$\mathbf{W}$, we run one forward/backward pass on cached activations from 2048 RedPajama
tokens (which are cached and used for repeated watermarking, e.g., using the secure inference server described in \Cref{sec:defenses}), and obtain the gradient $\mathbf{g}\!=\!\nabla_{\mathcal{I}} \mathcal{L}$.
We update $\Delta\mathbf{w}\leftarrow\Delta\mathbf{w}-\frac{\mathbf{g}^T
\Delta\mathbf{w}}{\lVert\mathbf{g}\rVert_2^2}\mathbf{g}$ to ensure
$\mathbf{g}^T\Delta\mathbf{w}=0$, eliminating first-order interference with our (proxy)
fine-tuning task.

\subsection{Decoding}
At verification time we read $\mathbf{W}$, compute correlations
$ r_k=\frac{1}{\gamma R}\sum_{j\in\mathcal{I}} W_j S_{k j}$, threshold at
zero to obtain hard bits $\hat{\mathbf{c}}$, and decode with the BCH decoder
$\mathcal{D}$; attribution succeeds iff
$\mathcal{D}(\hat{\mathbf{c}})=\mathbf{m}$.

\section{Other compression methods}
We also explore keeping the full precision of the top 1 percent of the weights with the highest magnitude, working from EasyQuant \citep{tang2024easyquant}. 
This uses the intuition that not all of the weights in a model will contribute equally to model performance.

Given a weight matrix $W \in \mathbb{R}^{m \times n}$, we identify the subset of weights with the highest magnitude as outliers:

\[
W_o = \{\, W_{ij} : \lvert W_{ij} \rvert \ge \tau \,\}
\]

where $\tau$ is the $(100-p)$-th percentile threshold of $|W|$ and $p$ is the outlier percentage at 1\%. The remaining weights are considered normal:
\[
W_n = \{\, W_{ij} : \lvert W_{ij} \rvert < \tau \,\}
\]

Unlike pruning approaches that remove parameters, this method preserves all weights but allocates precision non-uniformly. We maintain high precision for outliers while aggressively quantizing normal weights.





\cleardoublepage
\bibliography{full_bib}
\bibliographystyle{unsrtnat}

\end{document}